\def\mode{0} 
\if 0\mode
  \documentclass{article}
\usepackage[onehalfspacing]{setspace}  
\usepackage{amsthm,amsmath,amssymb}
\else
  \documentclass[ijoc,sglanonrev]{informs4}
\fi
\usepackage{secdot}
\usepackage{authblk}
\usepackage[square,sort&compress,comma,numbers]{natbib}

 \newcommand{\myqed}{\square}

\usepackage{tikz}
\usetikzlibrary{patterns,arrows.meta}

\usepackage{bbm}
\usepackage{amsmath}
\usepackage{graphicx}
\usepackage{natbib} 
\usepackage{xcolor}
\usepackage{float}
\usepackage[ruled,vlined,linesnumbered]{algorithm2e}
\usepackage{enumitem}
\usepackage{comment}
\usepackage{multirow}
\usepackage{cancel}
\usepackage{enumitem}
\usepackage{soul}
\usepackage[normalem]{ulem}
\usepackage{subcaption}
\usepackage{bbm}
\usepackage{wrapfig}
\usepackage[hyphens]{url}

\newcommand{\nourl}[1]{ }
\newcommand{\real}{\mathbb{R}}
\newcommand{\set}[2]{\left\{#1 \; \left|\;\; #2 \right.\right\}}
\newcommand{\abs}[1]{\left|#1\right|}
\newcommand{\card}[1]{\abs{#1}}

\newcommand{\Pmax}{\bar P}

\newcommand{\njobs}{m}
\newcommand{\nbins}{n}

\newcommand{\p}[1]{(\gamma_{#1}+\affine_{#1}x_{#1})_+}
\newcommand{\pu}[1]{\bar{p}_{#1}}
\newcommand{\ha}{\hat{a}}
\newcommand{\ba}{\bar{a}}
\newcommand{\UO}{\mathcal{U}_{\Omega}}

\newcommand{\tUO}{\tilde{\mathcal{U}}_{\Omega}}

\newcommand{\Pfmax}{P_f^{\max}}
\newcommand{\Pmin}{P^{\min}}
\renewcommand{\Pmax}{P^{\max}}
\newcommand{\Papprox}{P^{\prime}}
\newcommand{\prounded}{p'}
\newcommand{\hprounded}{\hat{p}'}
\newcommand{\affine}{\beta}

\renewcommand{\mp}[1]{{\color{black}#1}}
\newcommand{\mptwo}[1]{{\color{black}#1}}
\renewcommand{\ng}[1]{{\color{black}#1}}
\newcommand{\ngrev}[1]{{\color{black}#1}}
\newcommand{\revtwo}[1]{{\color{black}#1}}

\newcommand{\DPmatrix}{\mptwo{Z}}

\newcommand{\MP}[1]{\todo[inline,color=blue!40]{\textbf{M:} #1}}
\renewcommand{\NG}[1]{\todo[inline,color=red!40]{\textbf{N:} #1}}

\DeclareMathOperator*{\argmax}{arg\,max}

\newtheorem{proposition}{Proposition}
\newtheorem{theorem}{Theorem}
\newtheorem{lemma}{Lemma}
\newtheorem{observation}{Observation}

\newtheorem{remark}{Remark}
\usepackage{natbib}
\bibpunct[, ]{(}{)}{,}{a}{}{,}%

\begin{document}
\def\myAbstract{We study a robust extensible bin packing problem, under a budget uncertainty model where item sizes lie in the intersection of a box with a one-norm ball. \mptwo{We design an algorithm based on the column-and-constraint generation framework}, which alternates between solving a master robust bin-packing problem with a finite uncertainty set and solving a separation problem. We show that the separation is strongly NP-hard given solutions to the continuous relaxation of the master problem. Then, focusing on the separation problem for the integer master problem, we show that this problem becomes a special case of the continuous convex knapsack problem, which is known to be weakly NP-hard. Next, we prove that our special case remains NP-hard. We develop a pseudo-polynomial dynamic program (DP) and a fully polynomial-time approximation scheme (FPTAS) for our special case whose running times match those of the binary knapsack problem. Finally, our computational study shows that the DP can be significantly more efficient in practice compared with solving the problem with specially ordered set (SOS) constraints using advanced mixed-integer \ngrev{programming} (MIP) solvers. Our experiments also demonstrate the application of our separation problem method to solving the robust extensible bin packing problem, including separating based on approximate master solutions in intermediate iterations. We further enhance the algorithm by proposing custom symmetry breaking inequalities, valid when bins have identical costs per unit of excess capacity. Finally, a case-study, based on real elective surgery data, demonstrates the potential advantage of our model compared with the actual schedule and optimal nominal schedules.}

\def\myKeywords{Robust bin packing, extensible bin packing, convex knapsack, dynamic programming, FPTAS.}

\if 0\mode

\title{Robust Extensible Bin Packing and 
 a Convex Knapsack Problem}

\author[1]{Noam Goldberg}
\author[2]{Michael Poss}
\author[3]{Yariv N. Marmor}
\affil[1]{Department of Industrial Engineering \& Management, Ben-Gurion University of the Negev, \texttt{goldnoam@bgu.ac.il}}
\affil[2]{LIRMM, University of Montpellier, CNRS, France, \texttt{michael.poss@lirmm.fr}}
\affil[3]{Department of Industrial Engineering and Management, Braude College of Engineering, Karmiel, \texttt{myariv@braude.ac.il}}

\maketitle

\begin{abstract}
\myAbstract
\\

\end{abstract}
\else
\EquationsNumberedThrough    


\MANUSCRIPTNO{}

\TITLE{Robust Extensible Bin Packing and Revisiting the Convex Knapsack Problem}
\ARTICLEAUTHORS{
\AUTHOR{Noam Goldberg} 
\AFF{Department of Industrial Engineering and Management, Ben-Gurion University, \EMAIL{goldnoam@bgu.ac.il}}

\AUTHOR{Michael Poss}
\AFF{LIRMM, University of Montpellier, CNRS, France, \EMAIL{michael.poss@lirmm.fr}}

\AUTHOR{Yariv Marmor}
\AFF{Department of Industrial Engineering and Management, Braude College of Engineering, \EMAIL{myariv@braude.ac.il}}}

\ABSTRACT{\myAbstract}

    \maketitle
\fi


\section{Introduction}

\mp{Consider the problem faced by an appointment scheduler who must determine the number of work shifts and assign\ng{ment} of 
appointment\ng{s} (items) 
to shifts
(bins). The duration \ng{(or size)} of each appointment 
\ng{can vary considerably and can only be crudely estimated} 
in advance. 
\ng{Thus,} we 
\ng{study} a robust variant of the problem, where 
the set of possible durations is assumed to 
\ng{be given by a particular polytope}, \ng{typically} constructed 
\ng{based on the} available historical data. Furthermore,  we consider the model where \ng{the shifts (bins) do not have a hard capacity but} excess time 
\ng{can} be charged if the 
\ng{total} duration 
assigned to a work-shift exceeds 
\ng{a soft bound on the capacity} by charging a cost proportional to the excess time. In line with the robust optimization literature, the overall objective is to minimize the cost of assigning the appointments to shifts in the worst-case scenario.}

\ng{Attending to the details of our problem, we are} given $\njobs$ appointments, where each appointment {$i\in[\njobs]\equiv\{1,\ldots,\njobs\}$} has a nominal duration of $\bar a_i$ time units, and this duration can deviate by at most $\ha_i$. In the uncertainty model under consideration the total deviation of appointment duration is bounded by a given $0<\Omega < \sum_{i\in [\njobs]}\hat a_i$. Accordingly, the uncertainty set considered for the item sizes is defined as
$$
\UO = \set{0\leq a \leq \hat a}{\sum_{i\in [\njobs]} a_i\leq \Omega}. 
$$
\ng{
\ngrev{This} uncertainty set has been} previously \ng{considered} in the robust optimization literature, for example, by~\cite{gounaris2013robust,OmerPossOJMO,tadayon2015algorithms} \ngrev{and it is also known as the ``continuous budgeted uncertainty set''~\cite[Page 51]{goerigk2024introduction}. The choice of $\Omega$ can be motivated by probabilistic bounds, for example the ones derived in~\cite{bertsimas2021probabilistic}.} \ngrev{In our problem, each work shift (bin) has an initial capacity of $V>0$, but overtime is allowed in each shift $j\in[\nbins]$, at the additional \ng{overtime} cost of $c_j$ per unit of time.} \ng{The cost of 
\ngrev{each} bin is the sum of a fixed cost for opening the bin and the overtime cost. 

Our problem of packing the items into a least cost set of bins is called \ngrev{robust extensible bin packing}~\eqref{eq:REBP}.} 
While deterministic (classical) bin packing was first used to model the assignment of elective surgeries to operating rooms~\citep{Dexter1999}, it is very important to address the uncertainty of surgery durations. At the same time, in this application typically the bin (workshift) capacities are not necessarily hard constraints. \ng{So, our problem} generalizes 
\ng{the classical bin packing problem \ng{in both of these important aspects, for which several studies have already appeared}} 
in the literature. First, it generalizes the robust bin packing with uncertain item sizes studied by~\cite{Bougeret2022,schepler2022solving}, where the 
overfilling of the bins is not allowed (equivalent to having $c_j=+\infty$ for all $j\in[\nbins]$ in~\eqref{eq:REBP}).  \cite{Bougeret2022} 
\ng{develop} 
constant-factor approximation algorithms for this problem. \cite{schepler2022solving} and~\cite{song2018robust} develop a branch-and-price algorithm for this problem. 
\ng{Second,} the current problem under consideration \eqref{eq:REBP} is {also} a robust generalization of the extensible bin packing problem, whose nominal (deterministic) version has been extensively studied, including  work on approximation algorithms~\citep{DellOlmo1998,coffman2006approximation,Levin2022}. Note that a problem similar to~\eqref{eq:REBP} has been introduced by~\cite{denton2010optimal}, also motivated by the allocation of surgery blocks to operating rooms, using a slight variation of the uncertainty set $\UO$, \mptwo{wherein it is the number of items that deviate that is bounded, instead of their total deviation as in $\UO$. Mathematically, the budgeted uncertainty set that they consider corresponds to normalizing each deviation in our definition of $\UO$ by its upper bound in the adversarial budget; see also~\cite[Page 51]{goerigk2024introduction}. \cite{denton2010optimal}} 
propose a conservative compact formulation providing an upper bound to the problem, as further detailed in~\cite{ardestani2021linearized}. \mptwo{\cite{denton2010optimal} also propose a two-stage stochastic model.}

{The contributions of this paper are as follows:} Our first contribution is to provide an exact \mptwo{(up to a tolerance parameter)} solution algorithm for~\eqref{eq:REBP} by means of a row-and-column generation algorithm, following the lines of~\cite{AyoubP16,zeng2013solving}, that 
\ngrev{alternates between} solving relaxed master problems and separation problems {that keep on generating scenarios to augment the relaxed master problem}. We prove that the separation problem is NP-hard in the strong sense given an arbitrary fractional solution of the continuous relaxation of the master problem. \mp{We further provide 
\ng{custom} valid inequalities to break the symmetry of the formulation, valid for the problem whenever $c_i=c_j
$ for all $i,j\in[n]$.} We then concentrate on integer solutions, in which case we show that the separation problem is \mp{a special case of the convex knapsack problem, where each convex function has only two pieces 
(the first one being flat) -- \ng{\emph{two-piece convex knapsack}}~\eqref{eq:CK}.  

We subsequently analyze the difficulty of solving 
\ng{our} special case of the convex knapsack problem.} First, we show that~{\eqref{eq:CK}} is NP-hard. Second, we develop a {fast} dynamic programming (DP) algorithm for~\eqref{eq:CK} that leverages the first-order optimality conditions of~\eqref{eq:CK} and reuses function evaluations between knapsack invocations (each for a different item being excluded). The resulting DP has a running time complexity bound in $O(n\Pmax)$\ng{, where $\Pmax$ is any upper bound on the maximum knapsack profit}, \mp{which compares favorably to the $O(n^{3/2}\Omega)$ prove\ng{n} 
for the 
general convex knapsack problem in~\cite{malaguti2019integer} whenever $\Pmax\in O( n^{1/2}\Omega)$}. \mp{Using the classical scaling and rounding technique,} our DP algorithm leads to an FPTAS of order $O(n^2/\epsilon)$, and we also show how an FPTAS for the binary knapsack problem may carry over to~\eqref{eq:CK}; \mp{we detail how both FPTASes compare favorably to the FPTASes derived for the more general convex knapsack problem (e.g. $O(\frac{n^4}{\epsilon^2})$ in~\cite{malaguti2019integer})}. We numerically illustrate our DP algorithm {as well as} a row-and-column generation algorithm \mp{with the proposed valid inequalities} on instances {used in} the knapsack and bin packing literature, respectively. The DP algorithm is compared with a type-two specially-ordered set (SOS-2) constrained formulation that is a sensible and common approach for solving the convex knapsack problems using general MIP solvers.

Finally, we consider an elective-surgery scheduling case study application for the robust bin packing problem algorithm that applies the fast DP for scenario generation. The case study is based on real elective surgery data of a medical department in a major hospital made available by a collaboration of SEE Lab (Technion, IIT) and a major Israeli hospital~\citep{SEELab2025}. Elective surgery scheduling is known as an important application of bin packing as noted above and considered by~\cite{Dexter1999} and extensible bin packing~\citep{denton2010optimal}, in particular. Mixed-integer programs and heuristics have been considered to solve stochastic and robust formulations with additional constraints beyond bin assignment in~\cite{freeman2016scenario} and~\cite{wang2024robust}, respectively.  The current case study demonstrates how patient data can be used to derive our bin packing instances and the advantage of solutions of our model compared with the actual schedule and optimal nominal schedules for 
\ngrev{three} different one week scheduling periods.

The rest of the paper is structured as follows. In the next section, we present the general solution algorithm for~\eqref{eq:REBP} and prove the strong hardness of the separation problem {in the case of fractional input vectors in Section~\ref{sec:fractional}}. In Section~\ref{sec:knapsack}, 
our complexity results and the DP algorithms for the convex knapsack problems \eqref{eq:CK}  are presented. 
Computational experiment results are then 
demonstrated in Section~\ref{sec:num}. \mp{Additional \ng{numerical} results concerning warmstarts 
\ng{and} proofs related to the FPTAS 
\ng{algorithms} are 
\ng{relegated to the appendix}.}

\section{Formulation and Solution Method}
\label{sec:formulation}

Given \ng{the item nominal size vector} $\ba$, 
\ng{the uncertainty set defining the possible deviations} $\UO$, and \ng{the bin capacity} $V$, our goal is to partition the 
\ng{$m$ items} into $n$ 
\ng{bins}, $B_1, B_2,\ldots,B_\nbins\subseteq [\njobs]$, $\bigcup_{j\in[\nbins]}B_j= [\njobs]$.
For each $B_j$, we have the bin cost function 
\[
f(B_j,a)=\mathbbm{1}\{B_j\neq\emptyset\}+c_j\left(\sum_{i \in B_j}\mptwo{(}\bar a_i+ a_i\mptwo{)}-V\right)_+,
\]
and we wish to solve the optimization problem
\begin{equation}
\tag{{REBP}}
\label{eq:REBP}
\min_{B_1, B_2,\ldots,B_\nbins\subseteq [\njobs] \atop \bigcup_{j\in[\nbins]}B_j= [\njobs]} \max_{a\in\UO} \sum_{j\in[\nbins]}f(B_j,a).
\end{equation} 
\mptwo{Figure~\ref{fig:example} illustrates 
\ngrev{the problem} by showing how the model \ngrev{may} favor solutions \ngrev{that} mix \ngrev{together} items with different $\hat a_i$ values.}


\begin{figure}
\centering
    \scalebox{0.85}{
\begin{tikzpicture}[font=\small]

\def\msc{0.018}            
\def\Vmin{230}             
\def\bw{1.05}
\def\lcol{black!40}        
\def\scol{black!10}        
\pgfmathsetmacro\Vy{\Vmin*\msc}
\pgfmathsetmacro\hbw{\bw/2}
\def\shA{5.0}
\def\shB{8.8}

\newcommand{\drawboxes}[5]{%
  \begin{scope}[shift={(#1,0)}]
    \pgfmathsetmacro\hone{(100+#4)*\msc}
    \pgfmathsetmacro\ttwo{(200+#4+#5)*\msc}
    \pgfmathsetmacro\alphaval{max(0,(200+#4+#5)-\Vmin)}
    \fill[#2] (-\hbw,0)     rectangle (\hbw,\hone);
    \fill[#3] (-\hbw,\hone) rectangle (\hbw,\ttwo);
    \pgfmathparse{\alphaval>0 ? 1 : 0}\ifnum\pgfmathresult=1
      \fill[pattern=north east lines, pattern color=black!85]
            (-\hbw,\Vy) rectangle (\hbw,\ttwo);
    \fi
    \draw (-\hbw,0) rectangle (\hbw,\ttwo);
    \draw (-\hbw,\hone) -- (\hbw,\hone);    
  \end{scope}}

\newcommand{\drawlabels}[7]{%
  \begin{scope}[shift={(#1,0)}]
    \pgfmathsetmacro\hone{(100+#5)*\msc}
    \pgfmathsetmacro\ttwo{(200+#5+#6)*\msc}
    \pgfmathsetmacro\alphaval{max(0,(200+#5+#6)-\Vmin)}
    \pgfmathsetmacro\overlab{(\Vy+\ttwo)/2}
    \pgfmathsetmacro\upy{ ((\hone<\Vy)*(\ttwo>\Vy)) ? (\hone+\Vy)/2 : (\hone+\ttwo)/2 }
    \node at (0,{\hone/2}) {$#2$};
    \node at (0,\upy)      {$#3$};
    \pgfmathparse{\alphaval>0 ? 1 : 0}\ifnum\pgfmathresult=1
      \node[fill=#4,inner sep=1pt,font=\smaller\footnotesize\bfseries] at (0,\overlab)
        {$\alpha_{#7}{\!=\!}\pgfmathprintnumber{\alphaval}$};
    \fi
    \draw (-\hbw,\ttwo) -- (\hbw,\ttwo);   
  \end{scope}}

\drawboxes{0.0}{\lcol}{\lcol}{45}{45}   \drawboxes{1.5}{\scol}{\scol}{0}{0}
\drawboxes{\shA+0.0}{\lcol}{\scol}{45}{0} \drawboxes{\shA+1.5}{\lcol}{\scol}{45}{0}
\drawboxes{\shB+0.0}{\lcol}{\scol}{50}{10} \drawboxes{\shB+1.5}{\lcol}{\scol}{30}{0}

\foreach \xa/\xb in {-0.62/2.12, \shA-0.62/\shA+2.12, \shB-0.62/\shB+2.12}{
  \draw[dashed,thick] (\xa,\Vy) -- (\xb,\Vy);
}
\node[left] at (-0.64,\Vy) {$V$};
\node[left] at (\shA-0.64,\Vy) {$V$};
\node[left] at (\shB-0.64,\Vy) {$V$};

\drawlabels{0.0}{\ell_1}{\ell_2}{\lcol}{45}{45}{1}   \drawlabels{1.5}{s_1}{s_2}{\scol}{0}{0}{2}
\drawlabels{\shA+0.0}{\ell_1}{s_1}{\scol}{45}{0}{1}  \drawlabels{\shA+1.5}{\ell_2}{s_2}{\scol}{45}{0}{2}
\drawlabels{\shB+0.0}{\ell_1}{s_1}{\scol}{50}{10}{1} \drawlabels{\shB+1.5}{\ell_2}{s_2}{\scol}{30}{0}{2}

\draw[-{Latex}] (-1.55,0) -- (-1.55,5.45) node[above]{\footnotesize Time (or cost)};

\node[below] at (0.0,-0.12)     {Bin 1};  \node[below] at (1.5,-0.12)     {Bin 2};
\node[below] at (\shA+0.0,-0.12){Bin 1};  \node[below] at (\shA+1.5,-0.12){Bin 2};
\node[below] at (\shB+0.0,-0.12){Bin 1};  \node[below] at (\shB+1.5,-0.12){Bin 2};

\pgfmathsetmacro\mixmid{(\shA+\shB)/2+0.75}
\node[align=center,font=\bfseries] at (0.75,5.92) {Grouped $\hat a$};
\node[align=center,font=\bfseries] at (\mixmid,5.92) {Mixed $\hat a$};
\node[align=center] at (0.75,5.52)      {\footnotesize$a{=}(45,45,0,0)$};
\node[align=center] at (\shA+0.75,5.52) {\footnotesize$a{=}(45,45,0,0)$};
\node[align=center] at (\shB+0.75,5.52) {\footnotesize$a{=}(50,30,10,0)$};

\pgfmathsetmacro\arl{2.30}
\pgfmathsetmacro\arr{\shA-0.80}
\pgfmathsetmacro\arm{(\arl+\arr)/2}
\draw[-{Latex[length=3mm]},very thick] (\arl,3.0) -- (\arr,3.0);
\node[above,font=\footnotesize] at (\arm,3.0) {re-pack};

\begin{scope}[shift={(-8,0.5)}]
  \fill[\lcol] (0,0.18) rectangle (0.30,0.44); \draw (0,0.18) rectangle (0.30,0.44);
  \node[right,font=\scriptsize] at (0.36,0.31) {$\ell$: large job, $\hat a{=}50$};
  \fill[\scol] (0,0.58) rectangle (0.3,0.84); \draw (0,0.58) rectangle (0.3,0.84);
  \node[right,font=\scriptsize] at 
  (0.36,0.68) {$s$: small job, $\hat a{=}10$};
  \fill[pattern=north east lines,pattern color=black!85] (0,-0.20) rectangle (0.30,0.06);
  \draw (0,-0.20) rectangle (0.30,0.06);
  \node[right,font=\scriptsize] at (0.36,-0.07) {overtime $\alpha_j$: part of box above $V$};
  \node[font=\scriptsize] at (2.5,-0.5)
    {box height $=\bar a_i+a_i$ in given scenario $a$
    };
\end{scope}

\end{tikzpicture}

    }
\caption{\ngrev{Consider an instance with four items $(\ell_1,\ell_2,s_1,s_2)$, where $\bar a_{\ell_i}=\bar a_{s_j}=100$, for $i,j\in[2]$. Also, $\hat a_{\ell_i}=50$ and $\hat a_{s_i}=10$ for $i\in[2]$, $c_j=1$ for $j\in [2]$, $V=230$ and $\Omega=90$. Mixing items $s_i$ and $\ell_i$ together in the right packing leads to a worst-case overtime cost $\theta=30$, instead of $\theta=60$ when these items are grouped as in the left packing.
\label{fig:example}}}
\end{figure}

Introducing a binary variable $y_j$, denoting whether bin $j\in[n]$ is opened, \mptwo{$\alpha_j(a)$ as a simple recourse slack variable corresponding to the deviation overfill of a bin under ``scenario'' $a$,} and variable $z_{ij}$, denoting whether item $i\in[m]$ is packed into bin $j\in[n]$, the mathematical program {for the robust extensible bin packing problem~\eqref{eq:REBP}} can be written as
\begin{subequations}
	\label{form:penOverflowBp}
	\begin{align}
		& \min_{y,z,\theta,\alpha} && \sum_{j\in [\nbins]}y_j+ \theta\\
		& \text{subject to} && \sum_{j\in[\nbins]}z_{ij}=1 && i\in [\njobs]\label{constr:assignconstr}\\
		& &&  z_{ij}\leq y_j && i\in[\njobs],j\in[\nbins]\label{constr:strong}\\
		& && \sum_{i\in[\njobs]}z_{ij}(\bar a_i + a_i) \leq Vy_j+\alpha_j(a) && j\in[\nbins], a\in \UO\label{constr:bincap}\\
		& &&  \sum_{j\in[\nbins]}c_j\alpha_j(a)\leq \theta && a\in \UO \label{cons:simpleRecourse}\\
		& && z\in\{0,1\}^{\njobs\times \nbins}, y\in\{0,1\}^n \nonumber\\
		& && \alpha(a), \theta \geq 0  && a\in \UO. \nonumber
	\end{align}
\end{subequations}
Here, constraint~\eqref{constr:assignconstr} ensures that every item is assigned to exactly one bin. Constraint~\eqref{constr:strong} ensures that every bin that is assigned an item must be open. 
Note that every method for solving this formulation must first address how to handle the infinitely many constraints~\eqref{cons:simpleRecourse} and~\eqref{constr:bincap}, together with the infinitely many variables $\alpha_j(a)$ for each $j\in[\nbins]$. 
We next describe the separation problem for solving~\eqref{form:penOverflowBp}.  Separating these constraints mean{s} 
given a solution $(y^*,z^*,\theta^*)$ of~\eqref{form:penOverflowBp}, determine which constraints~\eqref{cons:simpleRecourse} with  $\alpha_j(a)=(\sum_{i\in[\njobs]}z_{ij}(\ba_i+a_i)-V)_+$ are violated, if any.  {For this problem, as well as a generalization known as the convex knapsack problem, we describe computational techniques that are efficient in practice as well as approximation schemes in the next sections.}

The typical approach to handle the infinite number of variables and constraints of~\eqref{form:penOverflowBp} relies on scenario generation. The basic idea of that approach introduces two optimization problems related to~\eqref{form:penOverflowBp}. On the one hand, we consider the relaxed master problem, which is defined from~\eqref{form:penOverflowBp} by replacing all occurrences of $\UO$ with a given finite set of scenarios $\tUO \subset \UO$. 
On the other hand, given a solution $y^*,z^*,\theta^*$ to the relaxed master problem, the separation problem 
is to determine a vector $a\in \UO$ such that the associated system of constraints 
{cannot be satisfied} 
with $\alpha_j(a)\geq0$ for all $j\in [n]$,   
\begin{subequations}
\label{eq:constraints_to_be_separated}
\begin{align}
 	& && \sum_{i\in[\njobs]}z^*_{ij}(\bar a_i + a_i) \leq Vy^*_j+\alpha_j(a) && j\in[\nbins]\label{constr:eq:constraints_to_be_separated:1}\\
		& &&  \sum_{j\in[\nbins]}c_j\alpha_j(a)\leq \theta^*. \label{cons:eq:constraints_to_be_separated:2}
\end{align}
\end{subequations}
 { 
 \begin{proposition}
 \label{prop:sep}
 Given 
 $y^*,z^*,\theta^*$ that is feasible for~\eqref{form:penOverflowBp}, or its LP relaxation, there exists a violated constraint of~\eqref{eq:constraints_to_be_separated}, for some $a\in \UO$, if and only if 
 \begin{equation}
 \tag{{SEP}}
\label{eq:SEP}
\eta^*=\max_{a\in \UO} \sum_{j\in[\nbins]}c_j\left( \sum_{i\in[\njobs]}z^*_{ij}(\bar a_i + a_i)-Vy_{\mptwo{j}}^*\right)_+\ngrev{>\theta^*.}
\end{equation}
\end{proposition}
\if 0\mode
  \begin{proof}
\else
  \begin{proof}{Proof.}
\fi
Given $a\in\UO$, the system of constraints~\eqref{eq:constraints_to_be_separated} cannot be satisfied with 
$\alpha_j(a)\geq 0$ for $j\in[n]$ if and only if it cannot be satisfied for
$
\alpha_j(a)=\left( \sum_{i\in[\njobs]}z^*_{ij}(\bar a_i + a_i)-Vy_{\mptwo{j}}^*\right)_+.
$
This choice of $\alpha_j(a)$ 
must satisfy~\eqref{constr:eq:constraints_to_be_separated:1}, 
and~\eqref{cons:eq:constraints_to_be_separated:2} 
is violated if and only if $\eta^*>\theta^*$.
\if 0\mode
\end{proof}
\else
  $\myqed$
  \end{proof}
\fi

The separation problem defined in Proposition~\ref{prop:sep} applies to solutions of~\eqref{form:penOverflowBp} as well as to its LP relaxation. Interestingly, whenever $z^*$ is binary, problem~\eqref{eq:SEP} turns out to be a special case of the following piecewise-linear convex knapsack problem having at most two linear segments, the first of which is constant,
\begin{equation}
\tag{{2PCK}}
\label{eq:CK}
\max_{x\in\real^\nbins}\set{\sum_{j\in[\nbins]} p_{j}(x)\equiv\p{j}}{\sum_{j\in[\nbins]} x_j\leq\Omega,\; 0\leq x \leq u},
\end{equation}
where we assume $\pu{j}\equiv p(u_j) > 0$ for each $j\in[n]$ 
(otherwise there exists an optimal solution satisfying $x_j=0$), and define $\Pmax\equiv \sum_{j\in[n]}\pu{j}$.}

\begin{observation}
	Given $z^*\in\{0,1\}^{\njobs\times\nbins}$, problem~\eqref{eq:SEP} is equivalent to~\eqref{eq:CK} with
	\begin{equation}
		\label{eq:kp_notations}
		{\gamma_j=c_j\bigg(\sum_{i\in [\njobs]:z^*_{ij}=1}\bar a_i-V\bigg),\quad \affine_j = c_j,}
		\quad\text{and}\quad u_j=\sum_{i\in[\njobs]:z^*_{ij}=1}\hat a_i, 
	\end{equation} \ngrev{for all $j\in [n]$.}
\end{observation}
\noindent\ngrev{Note that without loss of generality it is assumed that $\gamma_j\leq 0$ for all $j\in [n]$ as positive values can be treated as  constants, so that problem~\eqref{eq:CK} can be written equivalently by extracting the constant out of the maximization.}

\begin{algorithm}
\SetKwInOut{Input}{Input}
\SetKwInOut{Output}{Output}
	\caption{Row-and-column generation algorithm \label{algo:rcg}}
    	\Input{Instance of~\eqref{eq:REBP}}
        \Output{$y^*,z^*,\theta^*$}
		Initialize $\tUO=\{
        \mathbf{0}\}$, $\tau=\tau_0$\;
        \texttt{scenario\_generated} $\leftarrow$ {false}\;
        \Repeat {\emph{\texttt{scenario\_generated}} = \emph{false}}{Solve~\eqref{form:penOverflowBp} with $\UO$ replaced by $\tUO$ 
		{with optimality gap $\tau$ to} obtain $y^*,z^*, \theta^*$\;\label{step:master}
		Solve separation problem~\eqref{eq:SEP} to obtain optimal $a^*$ and $\eta^*$\; 
		\If{$\eta^* > \theta^*$}
		{\texttt{scenario\_generated} $\leftarrow$ {true}\; 
        $\tUO\leftarrow \tUO\cup\{a^*\}$\tcc*[r]{Add vars $\alpha_{j,a^*}$ and cons~\eqref{eq:constraints_to_be_separated} related to $a^*$ to~\eqref{form:penOverflowBp}}}
		\lElse{
		\textbf{if} $\tau = \tau_0$ \textbf{then}
		{$\tau = \tau_1$}}
        }
\end{algorithm}

The introduction of the relaxed master problem and the separation problem naturally leads to a scenario generation algorithm, also called row-and-column generation algorithm, detailed in Algorithm~\ref{algo:rcg}. 
{The algorithm essentially iterates between solving relaxed master problems and separation problems until the 
relaxed master problem becomes feasible. 
A \mptwo{modest practical}
innovation 
of our Algorithm~\ref{algo:rcg} 
is the use of two different optimality gaps, $\tau_0$ and $\tau_1$.} The 
algorithm 
may involve two possible bottlenecks. On the one hand, the size of the relaxed master problem grows as 
{more} scenario{s} are generated, leading to increasingly larger MILPs to be solved. On the other hand, the separation problem is typically a 
{computationally challenging} problem 
{in its own right,} so efficient 
{custom} algorithms 
 {can be most useful. To mitigate the first bottleneck, the MILP may be solved inexactly, with some loose optimality gap $\tau_0$, in intermediate iterations of the algorithm and only once no scenarios can be generated given $\tau_0$, the gap is tightened to require exact solutions of the MILP.} Regarding the second bottleneck, the complexity of the separation problem 
 depends on whether the considered solution of the master problem is integral or not. 
 {Fractional (non-integer) solutions need to be considered when} solving the LP relaxation {of the master problem}, which 
 can be useful 
for generating {an initial set} 
 of scenarios before {considering the integrality restrictions on $y$ and $z$.} {In the next section we describe how these fractional solutions may lead to 
 intractable 
 separation problems.}

\mptwo{We make next two observations regarding the formulation~\eqref{form:penOverflowBp} studied in this paper.

\begin{remark} 
\ngrev{A}n upper bound $W$ on the overfill $\alpha_j(a)$ could be added to our model to make it more realistic in some applications. Adding the bound would lead to two changes in Algorithm~\ref{algo:rcg}, detailed next. First, the feasibility of the assignment $z^*$ would be tested prior to solving~\eqref{eq:SEP}, by verifying, for each $j\in[n]$
\begin{equation}
\label{eq:SEP0}\tag{SEP0}
\sum_{i\in[\njobs]:z^*_{ij}=1}\bar a_i+\min\left(\Omega,\sum_{i\in[\njobs]:z^*_{ij}=1}\hat a_i\right)\leq V+W.
\end{equation}
If the inequality in~\eqref{eq:SEP0} is violated for some $j\in[n]$, the corresponding scenario $a\in\argmax\limits_{0\leq a\leq \hat a}\sum\limits_{i\in[\njobs]:z^*_{ij}=1}a_i$ is added to $\tUO$, and the algorithm would proceed with solving the updated~\eqref{form:penOverflowBp}. Otherwise, the algorithm proceeds with solving~\eqref{eq:SEP}, modifying \ngrev{the bounds $u_j$, for all $j\in [n]$, in} ~\eqref{eq:kp_notations} to $u'_j=\min\left(u_j,V+W-\sum_{i\in[\njobs]:z^*_{ij}=1}\bar a_i\right)$. 
\end{remark}

\begin{remark} Our algorithm is based on the assignment formulation for the bin-packing problem. While this formulation is known to be weak \ngrev{for the nominal bin packing formulation} compared to the pattern-based \ngrev{(solved using the branch-and-price method)} and arc-flow formulations, the latter two formulation\ngrev{s} extend less naturally to the robust counterpart. 
\ngrev{For the first pattern based formulation,} the pricing problem 
\ngrev{is} an extensible version of the multi-dimensional knapsack problem, a strongly NP-hard problem that 
\ngrev{would need to} be solved to 
optimality at every node of the branch-and-bound tree. 
\ngrev{With respect to the latter,} arc-flow formulations would lead to a set of flow variables for each scenario generated, and \ngrev{are} likely to lead to master problems of prohibitive sizes.
\end{remark}
}

\subsection{Symmetry-Breaking in the case of Equal Cost Bins}

Assignment-based formulations for bin packing may admit symmetric solutions that are equivalent in terms of solution objective value. When bins have equal deviation overfill costs, that is $c_j=c$ for all $j\in [n]$, then simple symmetry breaking constraints
\begin{equation}\label{eq:simplebinsymbr}
y_1\geq y_2 \geq \cdots \geq y_n,
\end{equation} may cut off (symmetric) solutions that are 
\ngrev{equivalent up to} re-indexing of the bins. These symmetry breaking inequalities are well-known in the context of  assignment-based formulations of standard bin packing. In the context of our problem they can be combined with the following inequality for each $j\in[n-1]$, 
\begin{equation}\label{eq:overtimevalidineq}
(1-y_{j+1})(\sum_{i\in [m]}\bar a_i+\Omega - jV)c\leq \theta.
\end{equation} The validity of this inequality is straightforward since for any optimal integer solution, \ngrev{$y_{j+1}=0$} 
implies that 
\ngrev{at most} $j$ bins are opened, in which case the overtime is bounded from below by the sum of item sizes after filling each of the $j$ bins to their full capacity $V$. 
The strengthening of formulation~\eqref{form:penOverflowBp} using inequalities~\eqref{eq:simplebinsymbr} together with~\eqref{eq:overtimevalidineq} 
\ngrev{is} evaluated in Section~\ref{sec:num}.  

\section{Complexity of the Fractional Separation Problem}
\label{sec:fractional}

We {now} prove 
that, 
{unless the} master {problem} solution is 
assumed to be integer, the separation problem is strongly NP-hard. 
 {This result implies that} generating scenarios 
 based on infeasible fractional master {problem solutions} 
 {may} not be possible in pseudo-polynomial time, unless $\mathcal P=\mathcal{NP}$. {Further, i}n view of this result, the following section will focus on the special case where $y^*$ and $z^*$ are integer 
 {in order to} 
 focus on developing a pseudo-polynomial time algorithm for this  problem.

Before introducing the hardness {result and its} proof, we 
{develop a} useful result related to the optimization {of a convex function} over 
$\UO$.
\begin{lemma}
\label{lem:rounding}
Assume that $\ha_i=1$ for each $i\in[\njobs]$, and that $\Omega$ is an integer not greater than $\njobs$. Let $g$ be a convex and non-decreasing function defined over $\UO$, and consider 
\ngrev{a} solution $a^*\in\UO$ \ngrev{that is optimal} to the optimization problem $\max_{a\in\UO}g(a)$. Then, one can round $a^*$ to an optimal solution 
$a'\in\set{a\in\{0,1\}^\njobs}{\sum_{i\in[\njobs]}a_i = \Omega}$ in polynomial-time.
\end{lemma}
\if 0\mode   \begin{proof} \else   \begin{proof}{Proof.} \fi
First notice that due to the assumptions on $\hat a$ and $\Omega$ and the fact that $g$ is non-decreasing (such that for all $x,x'$, where component-wise $x'\geq x$, $g(x')\geq g(x)$), we can assume that there exists an optimal $a^*$ such that  $\sum_{i\in [\njobs]}a^*_i=\Omega$. Let $I(a^*)$ denote the fractional components of $a^*$. Of course, if $I(a^*)=\emptyset$ then $a^*$ is integer and otherwise, by integrality of $\Omega$, $\card{I(a^*)}\geq 2$. For some $\ngrev{i',i''}\in I(a^*)$ such that $
\frac{\partial^+ g}{\partial a_{i'}^+}(a^*)\geq \frac{\partial^+ g}{\partial a_{i''}^+}(a^*)$, let  $\epsilon=\min(1-a^*_{i'},a^*_{i''})$, and define the direction $d$ to be given, for each $i\in[\njobs]$, by
\[
d_i = \begin{cases}
	\epsilon & i=i'\\
	-\epsilon & i=i''\\
	0 & \text{otherwise.}
	\end{cases}
\] 
	By convexity of $g$, its directional derivative at $a^*$ in direction $d$, $g'(a^*;d)=\lim_{t\rightarrow 0_+}\frac{g(a^*+td)-g(a^*)}{t}$ also exists (see~\cite[Theorem 7.37]{Beck2014}) and following the choice of $\ngrev{i'},i''$,
	\[
	g'(a^*;d)=	\epsilon\left[\frac{\partial^+ g}{\partial a_{i'}^+}(a^*)-\frac{\partial^+ g}{\partial a_{i''}^+}(a^*)\right]\geq 0.
	\] Finally, again by convexity of $g$, $g(a^*)+g'(a^*;d)\leq g(a^*+d)$. 
	It follows that $g(a^*+d)\geq g(a^*)$, hence $a^*\mptwo{+d}$ must be optimal, while $\ngrev{\card{I(a^*+d)}<\card{I(a^*)}}$. 
	Repeatedly applying above argument, if $a^*+d$ has a fractional component then $\card{I(a^*+d)}\geq 2$, and the same procedure can be applied up to 
    \ngrev{$m-1$} times to determine a binary optimal solution. 
\if 0\mode \end{proof}\else   $\myqed$ \end{proof}\fi
{Using this result we prove the following complexity result for the fractional separation problem.}
\begin{theorem}
Given $z^*\in[0,1]^{\njobs\times\nbins}$ and $y^*\in[0,1]^{\nbins}$, the separation problem~\eqref{eq:SEP} is strongly NP-hard.
\end{theorem}

\if 0\mode   \begin{proof} \else   \begin{proof}{Proof.} \fi
 The proof follows a reduction from the $K$-clique problem, which is to determine whether there exists a clique of size at least $K$ in $G{=(N,E)}$. Denote the edges incident to a node $j\in N$, by $\delta(j)\subseteq E$. 
{We consider a reduction to an instance of the separation problem creating a bin for each node of $G$ (so $N=[\nbins]$) and an item for each edge (so $E=[\njobs]$).} Consider a (fractional) solution that is feasible to the LP relaxation of~\eqref{form:penOverflowBp} for some $\tUO\subset \UO$,  given by 
  $y^*_j=1/2$, for each $j\in N$, and  $z^*_{ej}=1/2$, for each $e\in \delta(j)$, 
  and $z^*_{ej}=0$,  otherwise. 
   Also, let $\bar a_e=0$ and $\hat a_e = 1$ for each $e\in E$, 
   $c_j=1$ for each $j\in N$, 
   $V=\mptwo{1}$ and {let} $\Omega=K(K-1)/2$. {Then, p}roblem~\eqref{eq:SEP} becomes
\begin{equation}
\label{eq:maxNPhard}
\max_{a\in \UO} \sum_{j\in N}\left( \sum_{e\in \delta(j)}a_e/2-1/2\right)_+=\frac{1}{2}\max_{a\in \UO} \sum_{j\in N}\left( \sum_{e\in \delta(j)}a_e-1\right)_+.
\end{equation}
Denote the objective 
{in} the right-hand-side of~\eqref{eq:maxNPhard} 
{by} $g(a)=\sum_{j\in N}\left( \sum_{e\in \delta(j)}a_e-1\right)_+$ and let $a^*$ be a 
solution {that is optimal} to
this maximization problem. Applying Lemma~\ref{lem:rounding} to $a^*$, 
it can be rounded to {an integer} $a'\in\set{a\in\{0,1\}^\njobs}{\sum_{i\in[\njobs]}a_i=\Omega}$ {in polynomial running time}, for which $g(a')=g(a^*)$. Let us denote the support of $a'$ by $E'\subseteq E$ and also let $N(E')\subseteq N$ denote the set of nodes covered by $E'$. The{n,} 
\begin{align}
{g(a')=}\sum_{j\in N}\left( \sum_{e\in \delta(j)\cap E'}1-1\right)_+=\sum_{j\in N}\left( |\delta(j)\cap E'|-1\right)_+&=\sum_{j\in N(E')}( |\delta(j)\cap E'|-1)\\
&=2|E'|-|N(E')|.\label{eq:rewrittenobj}
\end{align}
Observe that $|E'|=\Omega=K(K-1)/2$ because of Lemma~\ref{lem:rounding}. 
\ngrev{Then, observing that}~\eqref{eq:rewrittenobj} can 
be written as 
$\ngrev{K(K-1)}-\card{N(E')}$, \ngrev{it follows that}
~\eqref{eq:rewrittenobj} is maximized 
when $|N(E')|$ is 
\ngrev{minimum (for a given set of edges).}  
In particular, 
${g(a')=K(K-1)}-K$ if and only if $N(E')$ is a $K$-clique in $G$. 
\if 0\mode \end{proof}\else   $\myqed$ \end{proof}\fi

\section{Convex Knapsack}
\label{sec:knapsack}

Focusing on the separation problem given an integral solution $(y^*,z^*)$, the subproblem to be solved 
is a special case of the well-studied (continuous) convex knapsack problem. 
We prove the hardness of 
our particular special case with convex piecewise linear functions, each having two linear segments. {Then}, we develop solution methods, including a pseudo-polynomial time DP algorithm and an FPTAS.

\subsection{Complexity of{~\eqref{eq:CK}}}
\label{sec:complexity}


For general convex functions and in particular quadratic $p_j$ a proof of the NP-hardness of the convex knapsack problem can be found in~\cite{Levi2014}. For discontinuous step functions, the NP-hardness has been proven in~\cite{Burke2008}. The analysis to follow in this section establishes that the problem is NP-hard even in {the special case~\eqref{eq:CK}.}

We first recall an important property that will be exploited below, which follows from the fact that the objective function of~\eqref{eq:CK} is convex, being a positive sum of convex functions. So, since~\eqref{eq:CK} always has a feasible solution, 
{it must have} an extreme point optimal solution~\citep[Theorem 32.3]{Rockafellar1970}. This fact and the form of such 
extreme point solutions of~\eqref{eq:CK} is 
summarized by the following {observation.}
\begin{observation}\label{obs:basicsolnknp}
There exists a solution $x^*$ that is optimal to~\eqref{eq:CK}, given by some $S^* \subseteq N$ for which $x^*_j=u_j$ for all $j\in S^*$,  and $f\in [\nbins]$ for which $0  \leq x^*_f < u_f$ and where $x^*_j=0$ for all $j\in [\nbins]\setminus(S^*\cup\{f\})$.
\end{observation}

{
\begin{theorem}
	\label{thm:compeqslopes}
	Two-piece convex knapsack problem~\eqref{eq:CK} is NP-hard even with 
    equal slopes $\beta_j$ for all $j\in [\nbins]$.
\end{theorem}
}
\if 0\mode   \begin{proof} \else   \begin{proof}{Proof.} \fi
 {Consider a subset sum instance with item sizes $w_1,\ldots,w_n$ and integer $M$. 
Consider a reduction to convex knapsack~\eqref{eq:CK} 
 with $\beta_j=2$ and $\gamma_j=-w_j$, 
 $u_j=w_j$, for all $j\in[\nbins]$, and $\Omega=M$. Using this reduction, we next show that the subset sum instance is a `yes'-instance if and only if \eqref{eq:CK} has an optimal objective value of exactly $M$.
	First note that 
	\begin{equation}\label{eq:objbound}
		\sum_{j\in[\nbins]}(2x_j-u_j)_+\leq \sum_{j\in [\nbins]}x_j\leq \Omega=M.
	\end{equation}
	Consider the case that $w_1,\ldots,w_n$, $M$ is a `yes' instance, implying that there is some $S\subseteq[\nbins]$ 
    satisfying $\sum_{j\in S}w_j=M$. Then, $x$ 
    defined so that $x_j=u_j$, for $j\in S$, and $x_j=0$, for $j\in [\nbins]\setminus S$,  satisfies $\sum_{j\in [\nbins]}x_j\leq M$ so it is feasible for~\eqref{eq:CK}. 
	Following the bound~\eqref{eq:objbound}, and the fact that $\sum_{j\in[\nbins]}(2x_j-u_j)\mptwo{_+}=\sum_{j\in S}\pu{j}=M$, this $x$ must also be optimal to~\eqref{eq:CK}. 
	
	Now consider a `no'-instance of subset sum. 
	By Observation~\ref{obs:basicsolnknp} it follows that there exists some $x^*$ that is an extreme point optimal solution of~\eqref{eq:CK}. 
	The fact that there is no $S\subseteq [\nbins]$ satisfying $\sum_{j\in S}u_j=M$, together with Observation~\ref{obs:basicsolnknp}, implies that there exists an $f\in[\nbins]$, such that $0 < x^*_{f} < u_{f}$, so that $(2x^*_{f}-u_{f})<x^*_{f}$.
	It then follows from~\eqref{eq:objbound} that $\sum_{j\in [\nbins]}(2x^*_j-u_j)_{\mptwo{_+}} < \sum_{j\in[\nbins]}x^*_j\leq M$, thereby establishing the claim.}
\if 0\mode \end{proof}\else   $\myqed$ \end{proof}\fi


\subsection{Knapsack Based DP}
\label{sec:dp}

We now develop a DP and an FPTAS for~\eqref{eq:CK} based on the following variant of the knapsack DP. Hence, we call $u_k$ and $\pu{k}$ the weight and profit of each item $k\in[n]$. For $f\in[\nbins]$, let $\zeta_{f}(P,k)$ denote the least weight binary knapsack subset of $[k]\setminus\{f\}$ that has a profit of at least $P$, which can be computed using the following DP recursion given item set $[k]$, 
\begin{align}
\label{eq:zetadef}
\zeta_{f}(P,k)=\begin{cases}
	0   & P=0   \text{ and } (k=1 \text{ or } k=2>f)\\
	u_{k} & 0 < P \leq \pu{k} \text{ and } (k=1 \text{ or } k=2>f)\\
	\infty & P > \pu{k} \text{ and } (k=1 \text{ or } k=2>f)\\
	   \zeta_f(P,k-1) & k=f {>1}\\
   \min\{\zeta_f(P-\pu{k},k-1)+u_k,\zeta_f(P,k-1)\} & 
   \text{otherwise.}
\end{cases}
\end{align}  
For convenience, let $\zeta_{n+1}(P,n)$ refer to the binary knapsack least weight item subset, corresponding to a solution of~\eqref{eq:CK} without any item $f\in[m]$ being fractional ($0<x_f<u_f$). Let $\Pmax$ be an upper bound on the profit of the convex knapsack problem~\eqref{eq:CK}, and define 
$$
\hat p_f(x_f) 
=\begin{cases} \p{f} & 0\leq x_f \leq u_f  \\-\infty & \text{otherwise}\end{cases},
$$
for each $f\in [\nbins]$. Then, the solution of the convex knapsack problem is given by
\begin{equation}\label{prob:dpprob}
P^*=\max_{\substack{f\in [\nbins+1],\\ P\in [\Pmax]\\\Delta\in[0,\Omega]}} \set{P+\hat p_f(\Delta)}{\zeta_{f}(P,n)+\Delta\leq \Omega} = 
{\max_{f\in[\nbins+1]}}\max_{{P\in[\Pmax]}}\left\{P+ \hat p_f(\Omega-\zeta_{f}(P,n))\right\}, 
\end{equation}
{where the first equality in~\eqref{prob:dpprob} {is immediate} from Observation~\ref{obs:basicsolnknp}. The second equality in~\eqref{prob:dpprob} follows from the fact that $\hat{p}_j(\cdot)$ are nondecreasing for each $j\in[\nbins]$ so there must exist an optimal solution where the knapsack capacity is exhausted. Overall, computing $P^*$ involves computing all elements of the tables $\zeta_f(P,k)$, which is done in $O(n^2\Pmax)$, and computing the maximum in the right-hand side of~\eqref{prob:dpprob} in $O(n^2\Pmax)$, leading to a total complexity of $O(n^2\Pmax)$.}    


{The following observation suggests that for each} $i\in[\nbins]$,  DP evaluations can be reused 
between calls {$\zeta_i(P,k)$} and $\zeta_j(P,k)$ for 
$i,j\geq k+1$.
\begin{observation}\label{obs:dpreuse}
	For \ngrev{a given $k\in [n-1]$ and} $P\in [\Pmax]$
    \ngrev{, for all} $i,j\geq k+1$, ${\zeta_i}(P,k)=\zeta_{j}(P,k)$.  
\end{observation}
This reuse of DP function evaluations may only contribute to a constant improvement (halving) in the overall running time complexity of solving~\eqref{prob:dpprob}. We next show that for the particular case of interest with convex piecewise linear knapsack with two pieces (the first piece having a zero slope) the reuse of DP evaluations can lead to a linear improvement in the running time complexity. This improvement is best possible 
{in the sense that} the running-time complexity is similar to that of the binary knapsack DP {(assuming $\Pmax$ dominates $\log(n)$).

The following lemma is {rather straightforward so it is stated without proof. It is also} similar to one appearing  in~\cite{Burke2008}.
\begin{lemma}\label{lem:deriv}
	Suppose that $x^*$ is an extreme point solution of~\eqref{eq:CK} with fractional item $f\in[\nbins]$ such that $0 < x^*_{f} < u_{f}$. Then, for all $j\in[\nbins]$ such that $x^*_j=u_j$,  $\beta_j\geq \beta_{f}.$
\end{lemma}
\noindent Note that in~\cite{Burke2008} this property of optimal solutions was only used to speed up their solution method in practice while in our case it is a key property that is used in order to improve the worst-case $O(n^2\Pmax)$ complexity of evaluating~\eqref{prob:dpprob}.
	
Assume the knapsack items are sorted and renumbered so that they satisfy
	\begin{equation}\label{derivord}
        \beta_1\geq\cdots\geq\beta_n,
	\end{equation}
where ties are broken such that if $\beta_j=\beta_i$ for some $i<j$ then $u_j\geq u_i$. Using the prior observation and Lemma~\ref{lem:deriv}, we develop a fast dynamic programming algorithm for~\eqref{eq:CK} that avoids 
recomputation using~\eqref{eq:zetadef} for each $f\in[n-1]$. This idea is summarized by
	\begin{equation}\label{eq:fastdp}
	{\max_{f=n+1,n,n-1,\ldots,1}}\max_{{P\in[\Pmax]}}\left\{P+ \hat p_f(\Omega-\zeta_{f}(P,f-1))\right\},
    \end{equation}
and detailed in Algorithm~\ref{algo:DP}. The algorithm uses the fact that the items are ordered  following a nonincreasing order of slopes, that is~\eqref{derivord} is satisfied. It proceeds with invoking the DP calls first without excluding any of the items. This first step also serves as a preprocessing step whose function evaluations are stored in order to be reused in the subsequent iterations. Next, in the main loop of the algorithm, items are excluded following a nondecreasing order of slopes, that is the reverse order of~\eqref{derivord}. For each excluded item $f$ only the DP table entries up to $f-1$ need to be considered.

\begin{algorithm}[H]
\SetKwInOut{Input}{Input}
\SetKwInOut{Output}{Output}
{
	\caption{Dynamic programming algorithm\label{algo:DP}}
		\Input{An instance of~\eqref{eq:CK} satisfying~\eqref{derivord} (otherwise sort the items to satisfy~\eqref{derivord})}
        $P'=0$\;
        Evaluate $\zeta_{n+1}(\Pmax,n)$ using~\eqref{eq:zetadef} and store in $\DPmatrix\in \real^{\Pmax\times n}$ so $\DPmatrix_{P,j}=\zeta_{n+1}(P,j)$ $\forall (P,j)\in [\Pmax]\times[n]$\label{DP:computetablefirst}\;
        \For{$f= n+1$ \KwTo $1$}{
        {
        \lFor{$P= 1$ \KwTo $\Pmax$}
        {
          $P'=\max(P',P+\hat p_f(\Omega-\DPmatrix_{P,f-1}))$
        }
        }
        }
        \Output{$P'$}}
\end{algorithm}

The next theorem establishes the correctness and running time of the algorithm. 
\begin{theorem} 
{Algorithm~\ref{algo:DP} outputs a solution that is optimal to~\eqref{eq:CK} in $O(n(\Pmax+\log n))$ time.}
\end{theorem}
\if 0\mode   \begin{proof} \else   \begin{proof}{Proof.} \fi 
	Consider evaluating $\zeta_f$ defined by \eqref{eq:zetadef} enumerating (and excluding) every possible ``fractional'' item $f$  according to the reverse order $n,n-1,\ldots,1$, each invocation with subset of items $[f-1]$. This method has an objective value $P' \leq P^*$, where the inequality holds because {$\zeta_f(P,f-1)\geq \zeta_f(P,n)$ (the right-hand side expression being a relaxation), so that $\hat p_f(\Omega - \zeta_f(P,f-1)) \leq \hat p_f(\Omega - \zeta_f(P,n))$ for each $P\in[\Pmax]$ and $f\in[n]$}.
    

Now, suppose that $x^*$ 
		{is an extreme point optimal solution of~\eqref{eq:CK} in the sense of Observation~\ref{obs:basicsolnknp}}. Further assume without losing generality for all $j\in[n]$, that $x^*_j>0$ implies that $p_j(x^*_j)> 0$. 
		Let 
		\[
		f^*=\max\left\{\arg\max_{i\in S'}\{u_i-x^*_i\}\right\}, 
		\] 
		where $S'\equiv \arg\min_{i\in [n]:x^*_i> 0}\{{\beta_i}\}$ ($f^*$ is the item having the smallest second segment slope of the items in the support of $x^*$, considering a fractional item if there is one, and otherwise breaking ties by selecting the highest index). 
		If $f^*=\max\set{i\in [\nbins]}{x^*_i>0}$, then trivially
		\begin{equation}\label{eq:knapsacklessthanf}
					  P^*=\max_{P\in[\Pmax]}\{P+\hat p_{f^*}(\Omega-\zeta_{f^*}(P,f^*-1))\} \leq P'.
		\end{equation}
		If $\card{S'}=1$ then following Lemma~\ref{lem:deriv} and Observation~\ref{obs:basicsolnknp}, $x^*_j\in\{0,u_j\}$ for all $j\in[\nbins]\setminus\{f^*\}$. Further, following the ordering~\eqref{derivord}, then $x^*_j=0$ for $j=f^*+1,\ldots,\nbins$.
		So, 
		\eqref{eq:knapsacklessthanf} holds. 
		
		If $x^*_{f^*}=u_{f^*}$, then following Lemma~\ref{lem:deriv}, the definition of $f^*$ and Observation~\ref{obs:basicsolnknp}, it must be that $x^*_j=u_j$ for all $j\in[\nbins]$ such that $x^*_j>0$. Further, following the definition of $f^*$, $x^*_j=0$ for all $j=f^*+1,\ldots,\nbins$. Consequently, again,~\eqref{eq:knapsacklessthanf} holds. 
		
		Otherwise, the case that remains to be handled has $\max\{\gamma_{f^*},-\gamma_{f^*}\} < x^*_{f^*} < u_f^*$, $\card{S'}\geq 2$ and $f^* < \max\set{j\in[\nbins]}{x^*_j>0}$. It follows from Observation~\ref{obs:basicsolnknp} that for all $j\neq f^*$, $x^*_j\in\{0,u_j\}$. Then, 
		$S'\setminus\{f^*\}\neq\emptyset$ and letting $i = \max(S'\setminus\{f^*\})$, it follows that $i>f^*$, $\beta_i=\beta_{f^*}$. 
		Since $x^*_i>0$, it implies that $x^*_i=u_i$.  
		So defining 
	    $x'$ by \[
		x'_k=\begin{cases} u_k- (u_{f^*}-x^*_{f^*})  & k=i\\
			x^*_k + (u_{f^*}-x^*_{f^*})  = u_{f^*} & k=f^*\\
			x^*_k & \text{otherwise, }
		\end{cases}
		\] 
		where it can be observed that 
		the ordering~\eqref{derivord} implies that $x'_i=u_i- (u_{f^*}-x^*_{f^*})\geq x^*_{f^*}\geq 0$, so $x'$ is feasible, having the same objective value as $x^*$ and for all $j < i$, $x'_j\in\{0,u_j\}$. 
		Hence,~\eqref{eq:knapsacklessthanf} holds with $f^*$ replaced by $i$.
So, it follows that $P' = P^*$.

	Next, to analyze the computational complexity of this method note that 	
	 once $\zeta_{n+1}(P,n)$ is evaluated for $P\in[\Pmax]$, then by Observation~\ref{obs:dpreuse}, 
	for each $f\leq n$, and $k\leq f-1$, $\zeta_{f}(P,k)=\zeta_{n+1}(P,k)$, the latter of which has already been computed (in order to evaluate $\zeta_n(P,n)$). 
First, sorting the items {in a preprocessing phase of the algorithm} is $O(n\log n)$. The first invocation of the DP {at step~\ref{DP:computetablefirst}} of the algorithm, with $f=n+1$, is done in $O(n \Pmax)$ time. Each of the subsequent $n\Pmax$ invocations 
is  $O(1)$, thereby establishing the claimed running time complexity.
\if 0\mode \end{proof}\else   $\myqed$ \end{proof}\fi

\subsection{Fully polynomial-time approximation schemes}

In this section we discuss two different approaches for obtaining an FPTAS for~\eqref{eq:CK}. The first approach applies the DP from the previous section together with the classical scaling and rounding scheme. The second one leverages Observation~\ref{obs:basicsolnknp} to adapt FPTASes of the binary knapsack problem to~\eqref{eq:CK}. FPTASes that make use of DP tables carry over more efficiently following this approach.

Let us discuss first the extension of classical bounds to~\eqref{eq:CK}. Suppose the items are indexed according to the profit density ordering 
\[
\frac{{\pu{1}}}{u_1} \geq \frac{{\pu{2}}}{u_2} \geq \cdots \geq \frac{{\pu{n}}}{u_n}.
\]
Now consider the profit lower bound, which is known for the knapsack problem, 
\begin{equation}\label{eq:knaplb}
\Pmin=\max\left\{\max_{j\in[\nbins]}\pu{j},{P_{{\text{R}}}}\right\},
\end{equation} where $P_{{\text{R}}}=\max_{k\in[\nbins]} \set{\sum_{j\in[k]}\pu{j}}{\sum_{j\in[k]}u_j\leq\Omega}$.
\begin{algorithm}
\SetKwInOut{Input}{Input}
\SetKwInOut{Output}{Output}
	\caption{Scaling and rounding algorithm for two-piece convex knapsack\label{alg:scalingalg}}
		{\Input{An instance of~\eqref{eq:CK}}}
        \Output{$P'$}
		Compute bounds 
		$\Pmin$,  $\Pmax$\;
		
		Define the scaled profits $\prounded_j = \lfloor\frac{\pu{j}}{K}\rfloor$ where $K=\frac{\epsilon \Pmin}{n}$\;
		Define $\Papprox=K\cdot \tilde P$, where $\tilde P$ is the solution of Algorithm~\ref{algo:DP} for the instance with scaled profit vector $\prounded$, $\Pmax$ replaced by $\lfloor\Pmax/K\rfloor$, and $\hprounded_k(\cdot)=
		\hat p_k(\cdot)/K
		$\;
\end{algorithm}

We establish the FPTAS approximation result using Algorithm~\ref{alg:scalingalg}.
\begin{proposition}\label{prop:improvedcomplex}
	For $\epsilon>0$, Algorithm~\ref{alg:scalingalg} outputs a $(1-\epsilon)$-approximation of~\eqref{eq:CK} in $O(n^2\epsilon^{-1})$ time.  
\end{proposition}
\begin{proof}
See appendix~\ref{app:FPTASscaling}.
\end{proof}

The following proposition presents an alternative approach that extends any \ngrev{binary} knapsack FPTAS for computing $\zeta_f(P,k)$ to an FPTAS for~\eqref{eq:CK}. 
\begin{proposition}~\label{prop:sepcomplex}
	\begin{enumerate}[label=(\roman*)]
	\item\label{propitem:fptasext1} Let $P\in[\Pmax]$ and let $g(\nbins,\epsilon)$ be the running time of any knapsack approximation algorithm that outputs
	$\zeta_{
		n+1}(\lceil P(1-\epsilon)\rceil,n)$.   Then, determining a solution of~\eqref{eq:CK} with objective at least $(1-\epsilon)P^*$ takes $O(n\log \Pmax g(n,\epsilon))$ time.
	\item\label{propitem:fptasext2} Let 
	$h(\nbins,\epsilon)$ be the running time of the knapsack FPTAS that outputs 
	$\zeta_{n+1}(\lceil \Pmax(1-\epsilon)^k\rceil,n)$ for all $k\in  I\equiv\{\lfloor-\log_{1-\epsilon}(\Pmin)\rfloor,\ldots,\lceil-\log_{1-\epsilon}(\Pmax)\rceil\}$ and further suppose that $h(n,\epsilon) {\in\Omega}(n\log n)$.
	 Then, determining a solution {for~\eqref{eq:CK}} with objective at least $(1-\epsilon){P^*}$ takes a total running time of $O(\nbins h(\nbins,\epsilon))$. 
	 \end{enumerate}
\end{proposition}
\begin{proof}
See appendix~\ref{app:FPTASother}
\end{proof}

Using the knapsack DP-based approximation of~\cite{Lawler1979}, for example, Proposition~\ref{prop:sepcomplex}-{\ref{propitem:fptasext2}} implies an overall running time complexity bound of {$O(n^2 \log(\epsilon^{-1}) + n \epsilon^{-4})$} for a convex knapsack FPTAS.
This is in contrast with the recent knapsack FPTAS~\citep{chen2024nearly} with running time $\tilde O(n+\epsilon^{-2})$, 
which may not be 
applicable for approximating convex knapsack using Proposition~\ref{prop:sepcomplex}{-\ref{propitem:fptasext2}} because it does not involve the computation of $\zeta_{f}(\lceil \Pmax(1-\epsilon)^k\rceil,n)$ for all $k\in I$.  
Following Proposition~\ref{prop:sepcomplex}-\ref{propitem:fptasext1} {instead,} the overall complexity bound for solving our problem is $\tilde O(n\log(\Pmax)(n+\epsilon^{-2}))$. 
Note that the approach of adapting knapsack methods based \ngrev{on} Proposition~\ref{prop:sepcomplex} 
applies more generally to \ngrev{the} convex knapsack problem. Finally, note that although applying our fast DP within a simple rounding and scaling approximation scheme is not dominated by the ``black box'' application of current knapsack FPTASes following the results of Proposition~\ref{prop:sepcomplex}, its complexity can be further improved using techniques such as those used in~\cite{Ibarra1975,Lawler1979}  including separately handling items that are considered small and more advanced proportionate scaling of larger items.

\section{Computational Results}
\label{sec:num}

We now evaluate the convex knapsack DP, Algorithm~\eqref{algo:DP}, and the robust bin packing row-and-column generation scheme,  Algorithm~\ref{algo:rcg}. \ngrev{All experiments were run on 
a linux server having an 
AMD 9655P CPU with 96 physical cores.} 
\ngrev{All experiments involving Gurobi use 
its default setting that restricts the maximum number of threads to 32 (our DP implementation is single-threaded)}. 
All of the algorithms were coded in Python and the main DP subroutines were pre-compiled using Numba. The mathematical programming models were coded in Python using the Pyomo package and solved using Gurobi version \ngrev{13.0.2}.

\subsection{Convex Knapsack - The Separation Problem}

First, we consider random instances of~\eqref{eq:CK}, and compare Algorithm~\ref{algo:DP} with 
a MIP on specially-ordered set (SOS) type-2 constraints~\citep{beale1970special}. \ngrev{Such constraints are a well established choice for modeling univariate piecewise-linear functions in optimization and especially when the number of breakpoints is small (as in our case) it is known to be a most efficient choice for such problems in state-of-the-art MIP solvers~\citep{de2013branch,rebennack2016computing,huchette2023nonconvex}.} 
A\ngrev{n SOS} 
constraint requires that a set of (continuous) variables satisfy a particular sparsity pattern, which can be handled  by a custom branching scheme within the branch-and-bound method. In particular, SOS-2 
{allows only two adjacent variables to be nonzero}, which is useful for representing piecewise linear functions as convex combinations of their breakpoints. {Letting $x_{j1}=0$, $x_{j2}=\min(
	(-\gamma_j/\beta_j)_+,u_j)$ 
    and $x_{j3}=u_j$, the resulting formulation for~\eqref{eq:CK} is
	\begin{subequations}
		\label{form:convknapsacksos}
		\begin{align}
			& \max_t && \sum_{j\in [\nbins]} {(\gamma_j+\affine_jx_{j1})_+}t_{j1}+{(\gamma_j+\affine_jx_{j2})_+}t_{j2}+{(\gamma_j+\affine_jx_{j3})}t_{j3}\\
			& \text{subject to} && \sum_{j\in [\nbins]}x_{j2}t_{j2} + u_jt_{j3} \leq \Omega\\
			& && t_{j1}+t_{j2}+t_{j3} = 1 && j\in [\nbins]\\
			& && t_{j1}, t_{j2}, t_{j3} \text{ is SOS-2,} && j\in [\nbins]
		\end{align}
	\end{subequations}
{where SOS-2 means that at most two consecutive variables of $(t_{j1},t_{j2},t_{j3})$ are non-zero.} Our method for generating hard to solve {instances of~\eqref{form:convknapsacksos} is based on} inversely correlated classical knapsack instances~\cite{Pisinger2005}: We 
{define $(\gamma_j+\affine_j x_j)_+=(\frac{x_j-\bar x_j}{u_j-\bar x_j}\pu{j})_+ $}}
where for each \ngrev{$j$}, $\pu{j}$ is generated uniformly at random from $(0,R)$ for values of $R=10^2,10^3,10^4$, and setting $u_j=\lceil\pu{j}+R/10\rceil$. 
{We also have the} additional parameter $\bar x_j$, for each $j\in [\nbins]$, which  is 
set to $\bar x_j=u_j-1$, which is found to maintain the challenge (of the associated randomly generated binary knapsack problem) in solving the 
\ngrev{\eqref{eq:CK}} instances to optimality using branch-and-bound solvers. For each value of $n$, 
\ngrev{thirty} experiments are run 
\ngrev{each with} knapsack capacity $\Omega=\frac{\ngrev{5+3i}}{101}\sum_{j\in N} u_j$ 
\ngrev{for} $i=1,\ldots,30$. \ngrev{We also compare with a weight-based version of the DP, with state space of dimension $\Omega$ in place of $\Pmax$, when using the inverse of $\zeta$ in Algorithm~\ref{algo:DP}. For completeness, the details of this algorithm are given in Apppendix~\ref{app:omdp}.} The results of these experiments are shown in Table~\ref{table:one}. As seen in the table the running time of 
\ngrev{both} DP \ngrev{variants} increases in $R$, as can be expected, but it still outperforms the solution of the SOS formulation~\eqref{form:convknapsacksos} using 
Gurobi 
(running in a single thread configuration), 
and of course is also much more stable (with a smaller variability as indicated by the half range).

\begin{table}[H]
	\begin{center}
\caption{Elapsed time in seconds for the SOS formulation~\eqref{form:convknapsacksos} solved using Gurobi vs. the DP method with \mptwo{$R=10^2,10^3,10^4$}. \ngrev{The second variant, $\Omega$-DP, is the corresponding weight-based variant of Algorithm~\ref{algo:DP}. The ``Avg w/l'' column displays the average running times including runs that 
\revtwo{reached} the 1800 sec. time limit.}
\label{table:one}}
\begin{tabular}{ccrrrrrrrrrrr}
 $\mathbf{n}$ & $\mathbf{R}$ & \multicolumn{4}{c}{\textbf{SOS}} & \multicolumn{3}{c}{\textbf{DP \ngrev{(Algorithm~\ref{algo:DP})}}} & \multicolumn{3}{c}{\ngrev{\textbf{$\Omega$-DP (Appendix~\ref{app:omdp})}}} \\
	& & Avg & Max & Avg w/l & Limit & Avg & Max & Limit  & Avg & Max & Limit\\
	\hline
	\hline
\multirow{2}{*}{100}   
&  $10^2$  &  0.52  &  6.61  &  0.52  &  0  &  0.03  &  0.05  &  0  &  0.04  &  0.08  &  0\\
&  $10^3$  &  31.79  &  799.80  &  31.79  &  0  &  0.25  &  0.46  &  0  &  0.35  &  0.74  &  0\\
&  $10^4$  &  9.75  &  239.23  &  129.11  &  2  &  2.76  &  5.18  &  0  &  4.11  &  9.00  &  0\\

\hline
\multirow{3}{*}{250}  
&  $10^2$  &  8.82  &  184.42  &  187.94  &  3  &  0.19  &  0.33  &  0  &  0.25  &  0.51  &  0\\
&  $10^3$  &  3.53  &  50.67  &  422.72  &  7  &  1.69  &  2.95  &  0  &  2.45  &  5.77  &  0\\
&  $10^4$  &  19.49  &  353.96  &  256.90  &  4  &  17.53  &  30.12  &  0  &  27.12  &  56.83  &  0\\
\hline
\multirow{3}{*}{500} 
&  $10^2$  &  31.89  &  757.16  &  326.58  &  5  &  0.71  &  1.46  &  0  &  0.93  &  2.07  &  0\\
&  $10^3$  &  83.58  &  1234.66  &  369.66  &  5  &  6.55  &  12.81  &  0  &  11.00  &  28.26  &  0\\
&  $10^4$  &  2.65  &  19.65  &  721.61  &  12  &  72.49  &  124.25  &  0  &  118.49  &  265.55  &  0
\end{tabular}
\end{center}
\end{table}

\subsection{Robust Extensible Bin Packing}

We now conduct experiments to evaluate Algorithm~\ref{algo:rcg} in solving~\eqref{form:penOverflowBp}, \mptwo{using the weight-based DP from Appendix~\ref{app:omdp} in the separation problem because of the fractional costs used in the instances}. We use data based on~\cite{song2018robust} for $\njobs=30,60,90$ and an instance with $\njobs=20$ created by taking the first 20 items of the one with $\njobs=30$. For all $i\in[\njobs]$, we set $\hat a_i=0.4\bar a_i$. Also, $V=\frac{1}{8}\sum_{i\in [\njobs]}(\bar a_i+\hat a_i)$, $\Omega=0.1\sum_{i\in [\njobs]}\hat a_i$. For all $i\in[\njobs]$, $c_i=c$, set as $c=1.5/V$.  \ngrev{Here, we set the number of bins based on the straightforward upper bound obtained for the case of no overtime and all jobs deviating as much as possible,  $\nbins=\lceil2(\sum_{i\in[\njobs]}\ba_i+\Omega)/V \rceil$.}

The experiments of Table~\ref{table:twoa} involve relatively easy instances, as $a_{\max}=\max_{i\in \ngrev{[m]}}\bar a_i=20$. 
Table~\ref{table:twob} shows the running time results (maintaining $\tau_0=0.2$) in solving the more challenging instances with larger $a_{\max}=100$. 

\begin{table}[h]
	\centering
	\caption{{Elapsed time seconds for Algorithm~\ref{algo:rcg} \ngrev{(\cite{song2018robust} data)} with  $a_{\max}=20$, and objective coefficient $c=\ngrev{1.5/V}$}.\label{table:twoa}}
	\begin{tabular}{rrrrrrrrrrr}
		$\mathbf{m}$ & $\mathbf{\tau_0}$ &  $\mathbf{\tau_1}$ & 
		\multicolumn{3}{c}{\textbf{Elapsed Time}} & \multicolumn{2}{c}{\textbf{Iterations}} &  \textbf{Limit}\\
		& & & Avg & Max & Avg w/lim & Avg & Max & \\
		\hline
		\hline
        \multirow{3}{*}{20} 
& 0.1  &  0.01  &  1190.4  &  2507.0  &  1791.3  &  91.6  &  121  &  1\\
& 0.2  &  0.01  &  1040.2  &  2277.9  &  1656.1  &  87.9  &  131  &  1\\
& 0.4  &  0.01  &  1196.2  &  2447.6  &  1796.6  &  96.8  &  170  &  1\\

        \hline
        \multirow{3}{*}{20} 
        & 0.1  &  0.05  &  11.58  &  31.53  &  11.58  &  22.10  &  32  &  0\\
        & 0.2  &  0.05  &  9.87  &  33.55  &  9.87  &  21.00  &  31  &  0\\
        & 0.4  &  0.05  &  11.75  &  40.70  &  11.75  &  24.20  &  38  &  0\\
        \hline
        
		\multirow{3}{*}{30} 
& 0.1  &  0.05  &  25.35  &  87.48  &  25.35  &  34.60  &  49  &  0\\
& 0.2  &  0.05  &  22.69  &  61.17  &  22.69  &  34.80  &  62  &  0\\
& 0.4  &  0.05  &  16.32  &  34.49  &  16.32  &  37.50  &  73  &  0\\
		\hline
		\multirow{3}{*}{60} 
& 0.1  &  0.05  &  255.69  &  1821.64  &  255.69  &  59.30  &  160  &  0\\
& 0.2  &  0.05  &  162.72  &  832.15  &  162.72  &  57.30  &  109  &  0\\
& 0.4  &  0.05  &  1433.96  &  4576.05  &  1433.96  &  232.10  &  521  &  0\\
		\hline
		\multirow{3}{*}{90} 
& 0.1  &  0.05  &  451.02  &  677.95  &  451.02  &  77.50  &  93  &  0\\
& 0.2  &  0.05  &  898.81  &  2073.63  &  898.81  &  98.00  &  114  &  0\\
& 0.4  &  0.05  &  1504.26  &  5757.40  &  2073.79  &  139.30  &  207  &  1
\end{tabular}\end{table}
{In Tables~\ref{table:twoa} and \ref{table:twob}, the 
instances experimented with are solved to {near-}optimality 
using Algorithm~\ref{algo:rcg}. In our implementation the branch-and-bound search is initialized in each iteration using the solution of the previous iteration. This is done  using the Gurobi variable VarHintVal attribute. Following preliminary experiments shown in the appendix, 
warm-starting the branch-and-bound search using a given solution in the subsequent iteration in this manner 
\ngrev{can improve over} the alternatives 
 of no warm starts or only constructing an initial feasible solution through the Gurobi Start attribute. In the following experiments, of Table~\ref{table:twoa}, we compare the values $\tau_0=0.1,0.2,0.4$ reflecting the required optimality gap in solving the master problem in the first phase of Algorithm~\ref{algo:rcg}. 
$\tau_0=0.2$ seemed to perform best overall.   
We also experimented with symmetry-breaking inequalities~\eqref{eq:simplebinsymbr}-\eqref{eq:overtimevalidineq}. These results are shown in Table~\ref{table:twob} on the harder instances. Symmetry breaking appears to contribute to solving these instances faster and in most cases our novel inequality~\eqref{eq:overtimevalidineq} contributes beyond the standard bin symmetry breaking inequalities.

\begin{table}[h]
	\centering
	\caption{Elapsed time seconds for Algorithm~\ref{algo:rcg} on harder benchmark instances \ngrev{(from~\cite{song2018robust})} with $a_{\max}=100$. A time limit of two hours is imposed and $\tau_0=0.2$.}\label{table:twob}
	\begin{tabular}{rrrrrrrrrr}
$\mathbf{m}$ & $\tau_1$ & Ineq.  
& \multicolumn{3}{c}{\textbf{Elapsed Time}} 
& \multicolumn{2}{c}{\textbf{Iterations}} &  \textbf{Limit}\\
 & & & Avg & Max & Avg w/lim & Avg & Max  & \\
\hline
\hline
\multirow{3}{*}{20} 
&  0.01  & None &  4064.1  &  5571.8  &  4691.3  &  148.1  &  217  &  2\\
&  0.01 &  \eqref{eq:simplebinsymbr}  &  1389.7  &  3514.7  &  1389.7  &  154.3  &  243  &  0\\
 & 0.01 & \eqref{eq:simplebinsymbr},\eqref{eq:overtimevalidineq} &  840.1  &  1514.9  &  840.1  &  135.5  &  202  &  0\\
\hline
\multirow{3}{*}{20} 
&  0.05  &  None & 17.57  &  36.62  &  17.57  &  29.20  &  40  &  0\\
&  0.05  & \eqref{eq:simplebinsymbr} & 2.77  &  6.63  &  2.77  &  20.10  &  34  &  0\\
&  0.05 & \eqref{eq:simplebinsymbr},\eqref{eq:overtimevalidineq} &  1.68  &  6.33  &  1.68  &  16.80  &  28  &  0\\
\hline
\multirow{3}{*}{30} 
&  0.05 & None  &  19.56  &  48.46  &  19.56  &  33.70  &  50  &  0\\
&  0.05 & \eqref{eq:simplebinsymbr} &  9.73  &  27.10  &  9.73  &  31.70  &  56  &  0\\
&  0.05  & \eqref{eq:simplebinsymbr},\eqref{eq:overtimevalidineq}  &  3.89  &  6.79  &  3.89  &  27.50  &  39  &  0\\
\hline
\multirow{3}{*}{60} 

&  0.05  & None & 264.52  &  451.83  &  264.52  &  66.10  &  80  &  0\\
&  0.05  & \eqref{eq:simplebinsymbr} &  108.52  &  442.89  &  108.52  &  60.00  &  116  &  0\\
&  0.05 & \eqref{eq:simplebinsymbr},\eqref{eq:overtimevalidineq} &  39.03  &  58.90  &  39.03  &  69.10  &  86  &  0\\
 \hline
\multirow{3}{*}{90} 
&  0.05 & None &  1916.10  &  5344.67  &  5614.55  &  132.10  &  161  &  7\\
&  0.05 & \eqref{eq:simplebinsymbr} &  349.05  &  1710.05  &  349.05  &  69.30  &  136  &  0\\
&  0.05 & \eqref{eq:simplebinsymbr},\eqref{eq:overtimevalidineq} &  103.56  &  377.89  &  103.56  &  79.50  &  162  &  0\\
\end{tabular}
\end{table}



\section{Surgery Scheduling using Hospital Department Data}

In this section we 
describe a case study applying our methods to surgery scheduling, to which end we first provide details of the data, outline the predictive models and parameters of the optimization model. Finally, we  
report on the obtained results.

\subsection{Real-world Data Description}
To demonstrate the model using real-world data, we integrated multiple datasets from a large hospital in northern Israel (January 2004 to November 2007). From patient admission and hospitalization records, we obtained information about whether patients were admitted through the emergency department or directly for hospitalization, as well as the hospital departments the patient passed through en route to the operating room. Background data such as age and gender were also collected.
By combining these datasets, we were able to calculate the duration of hospitalization prior to surgery and identify the department that referred the patient for surgical intervention. From the operating room database, we retrieved the surgical classification codes (ICD-9 codes~\cite{ICD9} – with potential multiple classification codes per patient surgical day) and used these to calculate the surgical duration and the number of classifications for each patient. Our focus was specifically on the 
\ngrev{Urological} operating room.
The explained variable of the model is surgical duration (from the moment of patient entry to the moment of patient exit).
The features introduced into the learning model encompassed comprehensive patient and procedural data. {For instance,} the gender distribution showed $37\%$ women and $63\%$ men, with a patient age averaging 45 years and a standard deviation of 26. 


\subsection{Prediction Models}
To predict surgical duration, we chose to use a regression tree model with 
5-fold cross-validation (allocating in each fold $20\%$ of the data for model testing). In the first model, we limited the number of observations in the leaves to be at least three times the square root of the number of training data points. In the second model, we constrained the number of observations in the leaves to be at least two times the square root of the training data points. 

The first model generated a regression tree with 12 leaves, having a regression coefficient of determination $R^2=0.228$ on the training data and $R^2=0.223$ on the test data (see Fig. \ref{fig:decision_tree_V1}). The second model 
is a more complex regression tree with 19 leaves, demonstrating a slightly improved performance with 
$R^2=0.253$ on the training data and $R^2=0.245$ on the test data (See Fig. \ref{fig:decision_tree_V2}).

\begin{figure}[H]
    \centering
     \begin{subfigure}[b]{0.5\textwidth}
     \includegraphics[height=1.8in]{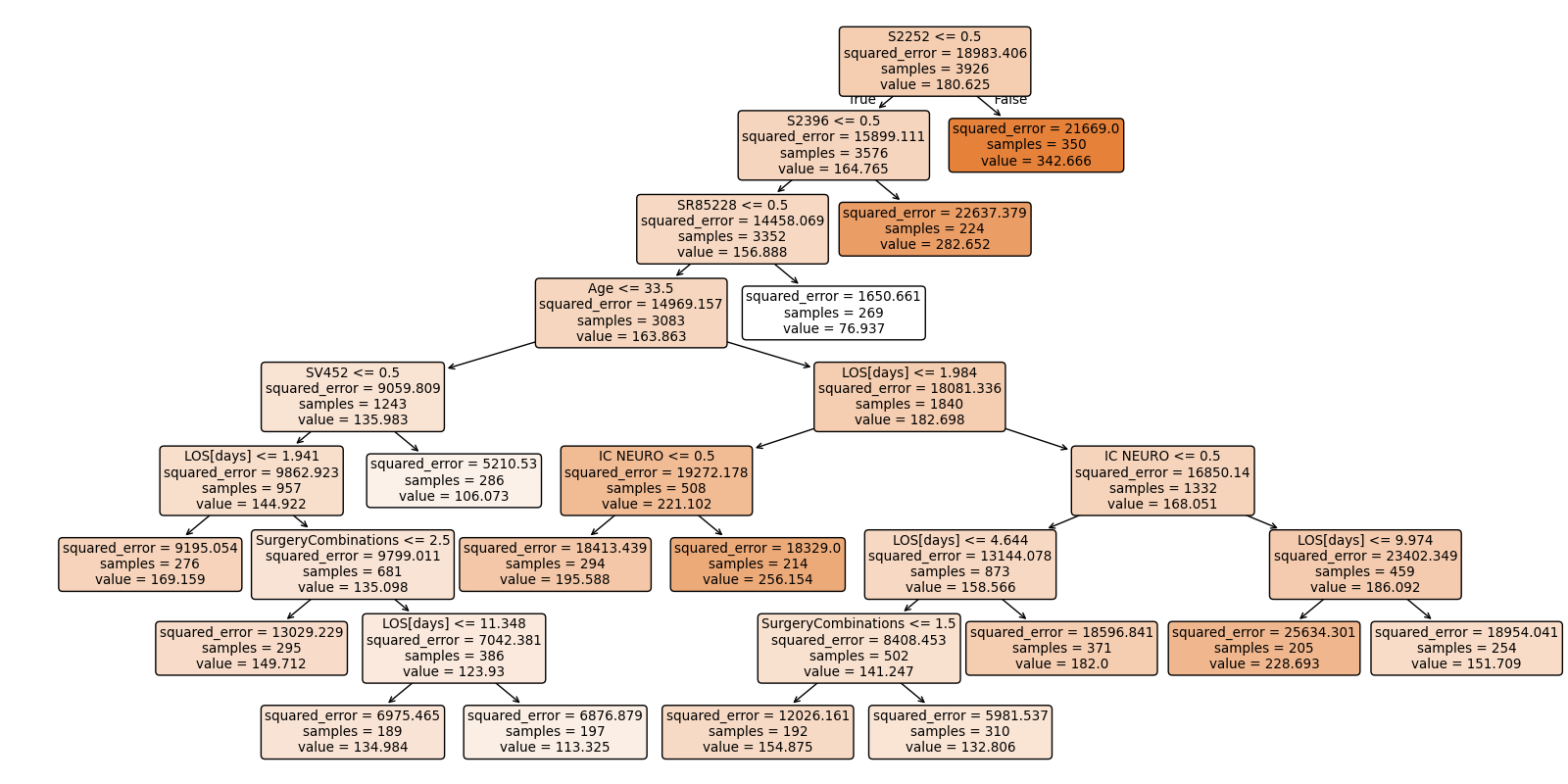}
    \caption{Small regression tree (V1) with $R^2\approx 0.22$ }
    \label{fig:decision_tree_V1}
    \end{subfigure}\begin{subfigure}[b]{0.5\textwidth}
    \includegraphics[height=1.8in]{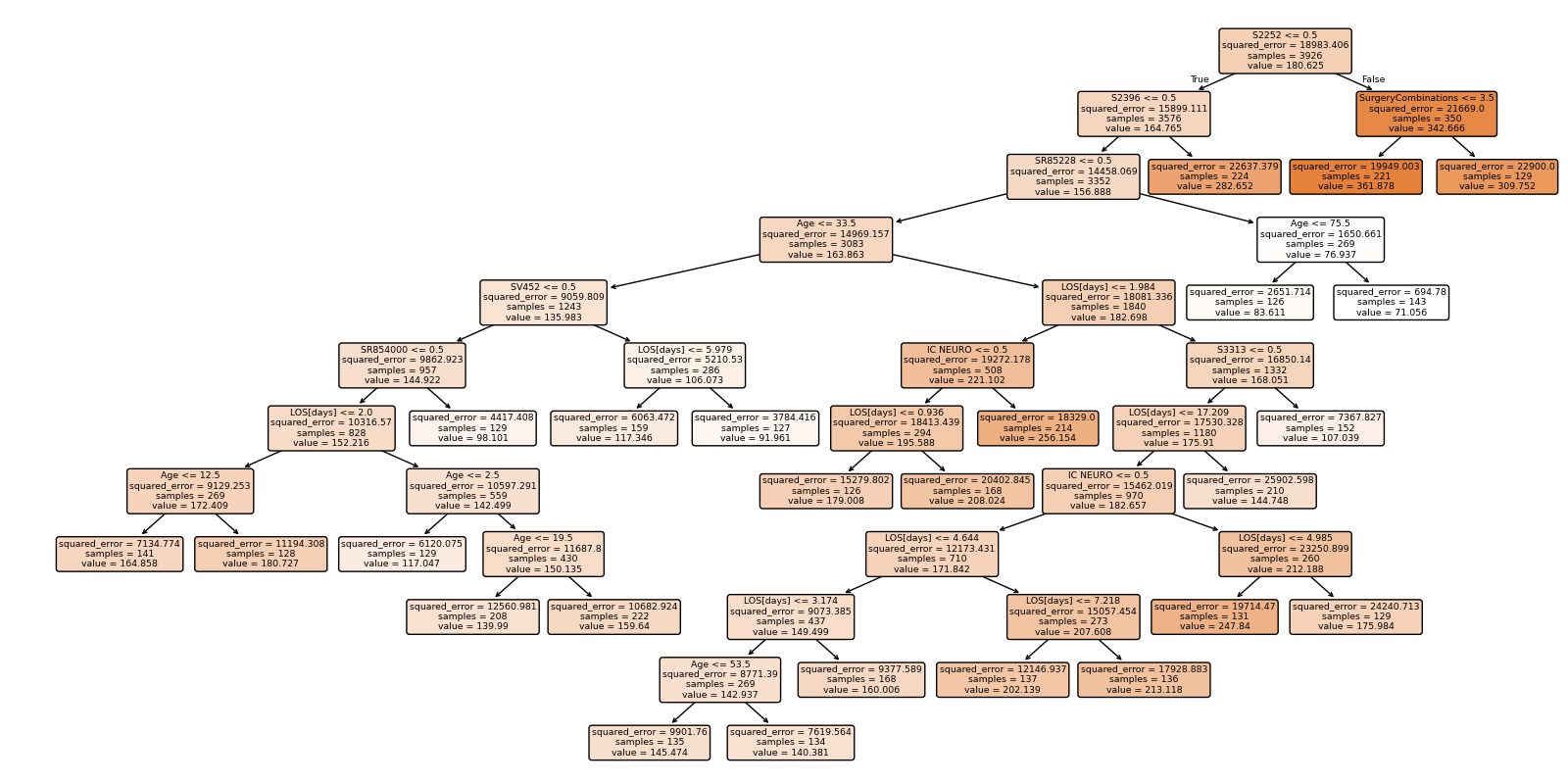}
    \caption{Large regression tree (V2) with $R^2\approx 0.25$}
    \label{fig:decision_tree_V2}
    \end{subfigure}
\end{figure}

\subsection{Robust Model Parameter Settings}

Applying our predictive modeling approach each regression tree leaf could be viewed as a surgery-patient category. A regression tree allows for a straightforward estimation of surgery duration distribution for each of these categories. Each patient $i$ is associated with a certain category (leaf) and accordingly it is associated with a nominal surgery duration $\bar a_i$ based on the mean or the predicted value at that leaf. Deviation $\hat a_i$ \ngrev{is} set according to the 95th percentile patient's category (leaf). $\Omega$ is estimated by examining the weekly distribution of the sum of surgery times, and setting it to the difference, subtracting the mean from the \ngrev{80th, 90th, or 95th percentile, resulting in $\Omega=477, 721, 971$, respectively. As is common in applying robust optimization to real data, using higher percentiles (or maximum) may lead to overly conservative solutions and poor performance, for example due to outliers; see Figure~\ref{fig:boxplot}.} 

\begin{wrapfigure}{r}{0.3\textwidth}
\centering
    \includegraphics[scale=0.25]{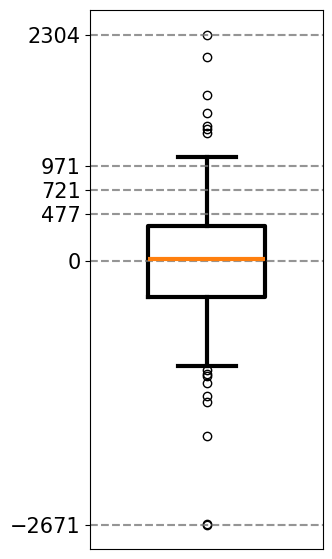}
    \caption{\ngrev{A box plot demonstrating the distribution of the total weekly surgery durations (normalized as differences from the mean). The 80th, 90th and 95th percentile values indicated are used to set $\Omega$.}}
    \label{fig:boxplot}
\end{wrapfigure}

\subsection{
\ngrev{Experiment Details}}

Initially, we examined which days of the week to investigate. We observed that on weekdays, there were between 2.8 and 3.1 surgeries on average per day, with a total average duration of 535 to 573 minutes daily. In contrast, weekend surgeries averaged between 1.6 and 2.2 per day, occupying only half of the total time. Consequently, we focused exclusively on regular workdays.
In the next stage, we analyzed which 12-week period (totaling 60 working days) was the most demanding in the year to establish it as the optimization model evaluation period. We discovered that from the second week of 2005 to the 13th week of 2005, the workload was significantly more intense. This was evident both in terms of the number of surgeries (17.66 average with 14.78 variance, compared to 14.87 average with 15.98 variance, p-value=0.015) and total surgery duration (3,352 weekly minutes versus 2,773 minutes, p-value=0.039). The period encompassed a total of 212 surgeries. From this dataset, we selected 
two representative weeks 
where the expected workload 
was relatively high \ngrev{(\revtwo{weeks 3 and 7} at 75\%-81\% \revtwo{average utilization})}, 
but not too high 
so that it could leave some room for improvement over the actual schedule, and another week with a relatively low workload \revtwo{(week 8 with 57\% utilization)}. In order to facilitate a meaningful comparison the workshift length is set as the minimal number of hours that would result in the nominal schedule booking all 5 days (to this end the overtime cost is set for all models to a relatively high value of $c=4/V$).

Table \ref{tab:utility_overtime} compares the robust model results 
with both the optimal nominal 
schedule (obtained by setting $\Omega=0$) 
and the actual schedule
. The results demonstrate that the robust model \ngrev{with the different values of $\Omega$ nearly always} outperforms both the nominal model and the actual implementation in terms of both utilization efficiency and overtime reduction. \ngrev{Moreover, the difference between the average daily performance and worst case (whether minimum utilization or maximum overtime) appears larger in most cases for the nominal model than for the robust models.} 

\begin{table}[H]
\centering
\caption{Utilization and Overtime Analysis}
\begin{tabular}{p{1.65cm}|p{2.2cm}|rrrr|rrrr}
\textbf{Period} & \textbf{Schedule Type ($\Omega$)} & \multicolumn{4}{c}{\textbf{Utilization}} & \multicolumn{4}{|c}{\textbf{Overtime}}\\
(work shift & & \multicolumn{2}{c}{\textbf{V1}} & \multicolumn{2}{c}{\textbf{V2}} & \multicolumn{2}{|c}{\textbf{V1}} & \multicolumn{2}{c}{\textbf{V2}} \\
 length) & & \textbf{Avg} & \textbf{Min} & \textbf{Avg} & \textbf{Min} & \textbf{Avg} & \textbf{Max} & \textbf{Avg} & \textbf{Max} \\
\hline
\hline
\multirow{5}{=}{Week 3\quad (8 hrs)} & Actual & 81\% & 24\%  &  81\% & 24\%  & 131 & 625 & 131 & 625 \\ 
& Nominal (0) & 90\% & 68\% & 96\% & 81\% & 87 & 410 & 61 & 200\\
& Robust (80\%) & 96\% & 86\% & 97\% & 86\% & 61 & 170 & 53 & 160\\
& Robust (90\%) & 95\% & 84\% & 97\% & 86\% & 62 & 310 & 53 & 160 \\
& Robust (95\%) & 95\% & 86\% & 97\% & 86\% & 62 & 310 & 53 & 125 \\
\hline

\multirow{5}{=}{Week 7 (10 hrs)} &  Actual & 75\% & 10\% & 75\% & 10\% &  66 & 210 & 66 & 210\\
&  Nominal ($0$) & 79\% & 48\%  & 79\%  & 44\% & 38   &  105  & 38 &  105\\
& Robust (80\%) & 83\% & 67\% & 86\% & 78\% & 16  & 80 & 0 & 0\\
&  Robust (90\%) & 82\% & 52\% & 86\% & 73\% & 22 & 105 & 1 & 5 \\
& Robust (95\%) & 86\% & 67\%  & 81\% & 38\% & 0 & 0 & 30 & 85\\
\hline
\multirow{5}{=}{Week 8\quad (7 hrs)} & Actual & 57\% & 10\% & 57\% & 10\% & 399 & 1640 & 399 & 1640\\
& Nominal (0) & 83\% & 57\% & 91\% & 76\% & 290 & 1225 & 254 & 1000\\
& Robust (80\%) & 90\% & 56\% & 91\% & 54\% & 260 & 1225 & 256 & 1075\\
& Robust (90\%) &  88\% & 57\% & 97\% & 93\% & 267 & 1230 & 230 & 1030\\
& Robust (95\%) & 82\% & 46\% &  89\% & 65\% &  294  & 1055 & 261 & 1205\\
\end{tabular}
\label{tab:utility_overtime}
\end{table}

\section{Conclusion}

In this paper we consider extensible bin packing with uncertain item sizes, which are assumed to be contained in a so-called budgeted uncertainty set. This models, for example, the scheduling and assignment of elective surgeries to surgery room shifts. The separation problem that generates scenarios from this uncertainty set is shown to be a special case of a two-piece convex knapsack problem. We show that this particular special case is NP-hard (in the ordinary sense). Further, given a master problem fractional solution, the separation problem is shown to be strongly NP-hard. Turning to algorithmic solution approaches for the (integer) separation problem, we develop a pseudo-polynomial DP algorithm and an FPTAS 
with improved complexity bounds. 
We numerically demonstrate the effectiveness of our DP algorithm for solving the two-piece convex problem, both compared to an SOS-based integer program, and specifically, for solving the robust extensible bin packing  separation problem.
Finally, we illustrate the benefit of considering the robust model for surgery scheduling 
using real data. Our results show that the schedule 
computed by the robust model can outperform both the actual schedules used in practice as well as 
the ones computed 
by solving the nominal model.




\revtwo{\section*{Acknowledgements}

We thank Stephen Shin for a prototype Python implementation that was extended for the computational work here. This work was partially supported by the MISTI MIT-Israel program. Noam Goldberg gratefully acknowledges related discussions with Jim Shi from NJIT regarding extensions of models in this paper as well as partial funding by a seed grant from the BGU–NJIT Institute for Future Technologies.

\section*{Data Availability Statement}
The code and bin-packing instances used in the current study are available in the GitHub repository \url{https://github.com/noamgold/robust-binpacking-convex-knapsack}.
}

\bibliographystyle{apalike-ejor} 
\bibliography{ref}

\if 0\mode
  \appendix
\else
  \begin{APPENDICES}
\fi

\section{$\Omega$-DP - the Weight-Based DP Variant\label{app:omdp}}

\mptwo{
We consider below the counterpart of dynamic programming recursion~\eqref{eq:zetadef} based on weights rather than profits. Accordingly, we define $\Pi_f(U,k)$ as the maximum profit
binary knapsack subset of $[k]\setminus \{f\}$ that has a weight of at most $U$, which can be computed using the following
\begin{align}
\label{eq:pidef}
\Pi_{f}(U,k)=\begin{cases}
	0   & U<u_k   \text{ and } (k=1 \text{ or } k=2>f)\\
	\pu{k} & u_k\leq U \text{ and } (k=1 \text{ or } k=2>f)\\
	   \Pi_f(U,k-1) & k=f >1\\
   \max\{\Pi_f(U-u_{k},k-1)+\pu{k},\Pi_f(U,k-1)\} & 
   \text{otherwise.}
\end{cases}
\end{align} 
Then, the solution of the convex knapsack problem is given by
\begin{equation}\label{prob:dpprobpi}
P^*= 
{\max_{f\in[\nbins+1]}}\max_{{U\in[\Omega]}}\left\{\Pi_f(U,n)+ \hat p_f(\Omega-U)\right\}.
\end{equation}
Computing $P^*$ through~\eqref{prob:dpprobpi} involves computing all elements of the tables $\Pi_f(U,k)$, leading to a total complexity of $O(n^2\Omega)$. Following the same approach as in Section~\ref{sec:dp}, we can instead compute $P'$ in $O(n\Omega)$, 
\ngrev{using} Algorithm~\ref{algo:DP2}. 

\begin{algorithm}[H]
\SetKwInOut{Input}{Input}
\SetKwInOut{Output}{Output}
{
	\caption{Dynamic programming algorithm based on the weights\label{algo:DP2}}
		\Input{An instance of~\eqref{eq:CK} satisfying~\eqref{derivord} (otherwise sort the items to satisfy~\eqref{derivord})}
        $P'=0$\;
        Evaluate $\Pi_{n+1}(
        \ngrev{\Omega},n)$ using~\eqref{eq:pidef} and store in $\DPmatrix\in \real^{\Omega\times n}$ so $\DPmatrix_{U,j}=\Pi_{n+1}(U,j)$ $\forall (U,j)\in [\Omega]\times[n]$\;
        \For{$f= n+1$ \KwTo $1$}{
        {
        \lFor{$U = 1$ \KwTo $\Omega$}
        {
          $P'=\max(P',\DPmatrix_{U,f-1}+\hat p_f(
          \ngrev{\Omega-U}))$
        }
        }
        }
        \Output{$P'$}}
\end{algorithm}
\ngrev{\noindent Following a similar analysis to that of} 
Section~\ref{sec:dp}, 
Algorithm~\ref{algo:DP2} \ngrev{outputs $P'=P^*$.} 
}

\section{Experiments with Warm-Starting Branch-and-Bound}\label{app:warmstarts}
We consider different strategies for warm-starting the branch-and-bound in line~\ref{step:master} of Algorithm~\ref{algo:rcg}. Warm-starting is performed by initializing the search with a feasible solution of the preceding iteration. This can be done either through the Start or VarHintVal attributes in Gurobi. Start only attempts to construct a feasible solution given some partial initialization of the variables (in our case the values of all the assignment variables $z_{ij}$ for \ngrev{$i\in [m]$ and $j\in [n]$}). The VarHintVal attribute affects the branch-and-bound search, for example node selection and branching strategy, beyond the determination of an initial feasible solution. The results of this experiment are shown in Table~\ref{table:algperformwarmstart}. 
\ng{Our} results \ng{suggest} that, \mptwo{for the most difficult instances}, initializing with the solution of the preceding iteration through \ngrev{either the Start or} the VarHintVal \ngrev{attributes} 
\ng{may} improve 
the \ngrev{succeeding} branch-and-bound search and overall algorithm performance, \mptwo{the impact being less clear for the easiest instances.} 
\ngrev{Accordingly, we have adopted the setting of the VarHintVal attribute throughout the rest of our experiments.}

\begin{table}[H]
	\centering
	\caption{Elapsed time seconds for Algorithm~\ref{algo:rcg} with  $a_{\max}=20$, and $\tau_0=\ngrev{0.2}$. 
    Warm Start indicates the method for passing the optimal solution of the preceding iteration to the branch-and-bound search of each iteration of Algorithm~\ref{algo:rcg}.\label{table:algperformwarmstart}}
	\begin{tabular}{rlrrrrrrrrr}
		$\mathbf{m}$ & \textbf{Warm Start} &  $\mathbf{\tau_1}$ & 
		\multicolumn{3}{c}{\textbf{Elapsed Time}} & \multicolumn{2}{c}{\textbf{Iterations}} &  \textbf{Limit}\\
		& & & Avg & Max & Avg w/limit & Avg & Max & \\
		\hline
		\hline
        \multirow{3}{*}{20} & \mptwo{None} & 0.01
        &  1378.8  &  3635.5  &  1378.8  &  93.6  &  148  &  0\\
& Start & 0.01 &  1316.7  &  3116.9  &  1905.0  &  93.4  &  134  &  1\\
 & VarHintVal & 0.01 & 1040.2  &  2277.9  &  1656.1  &  87.9  &  131  &  1\\
 \hline
        \multirow{3}{*}{90} & \mptwo{None} & 0.05 
        &  2735.5  &  4311.0  &  4074.7  &  237.3  &  358  &  3\\
        & Start & 0.05 & 471.5  &  770.3  &  471.5  &  80.6  &  95  &  0\\
        & VarHintVal & 0.05 &  898.8  &  2073.6  &  898.8 &  98.0  &  114  &  0\\
        \end{tabular}

\end{table}

\section{Proofs of the FPTASes}
\label{app:FPTAS}

\subsection{FPTAS from scaling and rounding}
\label{app:FPTASscaling}

\mp{Let us first prove that} the objective value of the general convex knapsack problem~\eqref{eq:CK} has bounds that are similar to those that are known for the binary knapsack problem, as shown by the following lemma.
\begin{lemma}\label{lem:bounds}
	$\Pmin\leq P^*\leq 2\Pmin$.
\end{lemma} 
\if 0\mode   \begin{proof} \else   \begin{proof}{Proof.} \fi
	The \ngrev{first} inequality is immediate from the fact that convex knapsack is a relaxation of the binary knapsack problem 
	so $
	P^*\geq P_{B}\geq \Pmin$, where $P_B$ is the optimal objective value of the binary knapsack problem instance with weights $u_j$, and profits $\pu{j}$ for $j\in [n]$.
    Observe that $P_{\text{LP}}\geq P^*$ (i.e., the LP relaxation of binary knapsack is also a relaxation of convex knapsack) \ngrev{-- it} follows from $(\gamma_j+\beta_j\lambda u_j)_+\leq \lambda \pu{j}$ \ngrev{(as it can be assumed without losing generality that $\gamma_j\leq 0$) for all $j\in [n]$ and $\lambda\in [0,1]$}. 
    So, $P_{\text{LP}}\geq P^*$. Finally, it is {straightforward} (it is also well known; see for example~\cite{Lawler1979}) that $2\Pmin\geq P_{\text{LP}}$.
    \if 0\mode \end{proof}\else   $\myqed$ \end{proof}\fi 

\mp{We are now ready to prove Proposition~\ref{prop:improvedcomplex}}.
\begin{proof}[Proof. of Proposition~\ref{prop:improvedcomplex}.]
	Letting $P^*$ denote the optimal convex knapsack objective and let $\Papprox$ denote the approximation algorithm objective value. Following Observation~\ref{obs:basicsolnknp},  considering optimal solution set of fully selected items $J^*$, let $f^*$ denote a fractional item or least index item in $[\nbins]\setminus J^*$ if there is no such item. Then, denote $z^*_{f^*}\equiv P^*-\hat p_{f^*}(\Omega-\sum_{j\in J^*}u_j) = \sum_{j\in J^*}\pu{j}$, and observe that
	\begin{equation}\label{eq:roundeq}
	z^*_{f^*}-K\sum_{j\in J^*}\prounded_j=\sum_{j\in J^*}(\pu{j}-K\prounded_j)\leq nK = \epsilon \Pmin.
	\end{equation}
	Then, letting $J'$ denote the subset of $[\nbins]\setminus\{f^*\}$ of fully used items that are output by the DP, evidently, 
	\begin{align*}
	\Papprox &\geq K\left(\sum_{j\in J'} \prounded_j+\hprounded_{f^*}\left(\Omega-\sum_{j\in J'}u_j\right)\right)\\
		&\geq 
		{	K\left(\sum_{j\in J^*} \prounded_j+\hprounded_{f^*}\left(\Omega-\sum_{j\in J^*}u_j\right)\right)}\\
		& \geq
		z^*_{f^*}+\hat p_f\left(\Omega-\sum_{j\in J^{*}}u_j\right)-\epsilon \Pmin\\
		&\geq (1-\epsilon)P^*,
\end{align*} 
{where the first inequality followed from the DP returning a solution that is optimal for the scaled profits and the second inequality followed from~\eqref{eq:roundeq}, and the fact that for all $x$ and $f$, $K\hat p'_f(x) = \hat p_f(x)$.} 
Since $\Pmax/K\leq n\Pmax/\Pmin\epsilon\leq 2n/\epsilon $, by Lemma~\ref{lem:bounds} the overall complexity is $O(n^2/\epsilon+n\log n)\subseteq O(n^2/\epsilon)$.
\if 0\mode \end{proof}\else   $\myqed$ \end{proof}\fi 

Proposition~\ref{prop:improvedcomplex} essentially shows that {applying the scaling and rounding scheme to problem~\eqref{eq:CK} leads to an FPTAS with a running time complexity bound similar to that of the binary knapsack problem}. {Note that while much more advanced FPTASes exist for the binary knapsack problem, some of which may extend to further improve on the running time of Algorithm~\ref{alg:scalingalg} (claimed by Proposition~\ref{prop:improvedcomplex}), it is not pursued herein to simplify the exposition.}

\subsection{Proof of Proposition~\ref{prop:sepcomplex}}
\label{app:FPTASother}

	\textbf{\ref{propitem:fptasext1}.}
    A $(1-\epsilon)$-approximate optimal value is computed by evaluating
	\[
	\Papprox = \max_{P\in [\Pmax]}\max_{f\in[\nbins]}\{\lceil P(1-\epsilon)\rceil+\hat p_f(\Omega-\zeta_f(\lceil P(1-\epsilon)\rceil,n))\}.
	\] {The outer maximization 
    determines the largest $P\in[\Pmax]$ such that $\hat p_f(\Omega-\zeta_f({\lceil P(1-\epsilon)\rceil},n))\geq 0$, which can be done using a binary search. This}
	results in $O(n\log\Pmax)$ invocations of the knapsack approximation algorithm. 
	
	\noindent\textbf{~\ref{propitem:fptasext2}.}
	Let $P^*_f$ denote the optimal objective value of the {(binary)} knapsack instance with item set $[\nbins]\setminus\{f\}$, {namely $P^*_f=\max\set{P}{\zeta_f(P,n)\leq \Omega}$ for $f\in[\nbins]$.} 
    Bounds $\Pfmax$ can be precomputed for all $f\in[n]$ in a total of $O(n\log n)$ time. 
	{Then, invoking the knapsack FPTAS, which by the hypothesis outputs an $\epsilon$-approximate solution for each $P\in[\Pmin_f,\Pmax_f]$ as a subroutine for each excluded item $f$, we get}
	\begin{align*}
	\Papprox &=\max_{f\in[\nbins]}\max_{k\in I}\{\lceil\Pfmax(1-\epsilon)^k\rceil+\hat p_f(\Omega-\zeta_f(\lceil\Pfmax(1-\epsilon)^k\rceil,n))\}\\ 
	&\geq {\max_{f\in[\nbins]}\{(1-\epsilon)P^*_f+\hat p_f(\Omega-\zeta_{{f}}(P^*_f,n))\}}\\
	&{\geq (1-\epsilon){\max_{f\in[\nbins]}\{P^*_f+\hat p_f(\Omega-\zeta_{{f}}(P^*_f,n))\}}}\\
	&={(1-\epsilon)P^*.} 
	\end{align*} 	
	The {first} inequality followed from the knapsack approximation for each $f\in[n]$, {specifically, that there exists $k^*\in I$ such that $(1-\epsilon)P^*_f\leq \lceil(1-\epsilon)^{k^*}\Pfmax\rceil\leq P^*_f$ (and from $\hat p_f(\cdot)$ {and $\zeta_f(\cdot,n))$} being nondecreasing}). The 
    \ngrev{final} equality followed from 
	the existence of such $f\in[\nbins]$, in particular for item $f$ that is fractional or 
	{zero} in the solution corresponding to $P^*$, by Observation~\ref{obs:basicsolnknp}. 

\if 0\mode
\else
  \end{APPENDICES}
\fi

\end{document}